\documentclass[a4paper,11pt]{article}
\usepackage{jheppub}
\usepackage[utf8]{inputenc}

\usepackage{lscape}

\usepackage{comment}

\usepackage{soul}

\usepackage{dsfont}
\usepackage{simplewick}

\usepackage{physics}
\usepackage{mathtools,amsmath,amssymb}

\DeclareMathSizes{1}{1}{1}{1}
\usepackage{exscale}

\usepackage{graphicx}

\usepackage{lmodern}
\usepackage[T1]{fontenc}
\usepackage{microtype}

\usepackage{xspace}

\usepackage{array} 

\usepackage{subfigure}

\usepackage{pgf}
\usepackage{pgfplots}
\pgfplotsset{compat=1.14}
\usepackage{tikz}
\usetikzlibrary{shapes,arrows,automata,positioning,backgrounds,intersections}
\usepackage{tikz-3dplot}

\usepgfplotslibrary{external} 

\newcommand{\YM}{{\mathrm{\scriptscriptstyle YM}}}

\usepackage{xspace}

 \usepackage{breqn}
  
\usepackage{placeins}

\DeclareMathOperator{\phaneq}{\phantom{{}=}}
\DeclareMathOperator{\sym}{sym}
\DeclareMathOperator{\anti}{anti}
\DeclareMathOperator{\sgn}{sgn}

\DeclareMathOperator{\idm}{\mathds{1}}

\usepackage[textsize=footnotesize
]{todonotes}

\makeatletter
\renewcommand{\todo}[2][]{\tikzexternaldisable\@todo[#1]{#2}\tikzexternalenable}
\makeatother

\makeatletter
\DeclareRobustCommand*{\bfseries}{%
  \not@math@alphabet\bfseries\mathbf
  \fontseries\bfdefault\selectfont
  \boldmath
}

\makeatother

\usepackage{lipsum}
\makeatletter
\if@titlepage
  \renewenvironment{abstract}{%
      \titlepage
      \null\vfil
      \@beginparpenalty\@lowpenalty
      \begin{center}%
        \bfseries \abstractname
        \@endparpenalty\@M
      \end{center}}%
     {\par\vfil\null\endtitlepage}
\else
  \renewenvironment{abstract}{%
      \if@twocolumn
        \section*{\abstractname}%
      \else
        \small
        \begin{center}%
          {\bfseries \abstractname\vspace{-.5em}\vspace{\z@}}%
        \end{center}%
        \quotation
      \fi}
      {\if@twocolumn\else\endquotation\fi}
\fi
\makeatother

\title{\texorpdfstring{From Hagedorn to Lee-Yang: Partition functions of $\mathcal{N}=4$ SYM theory at finite $N$}{From Hagedorn to Lee-Yang: Partition functions of N=4 SYM theory at finite N}}
\author{Alexander T. Kristensson and Matthias Wilhelm}

\begin{document}

\begingroup\parindent0pt
\begin{flushright}\footnotesize
\end{flushright}
\vspace*{4em}
\centering
\begingroup\LARGE
\bf
From Hagedorn to Lee-Yang: Partition functions of $\mathcal{N}=4$ SYM theory at finite $N$
\par\endgroup
\vspace{2.5em}
\begingroup\large{\bf Alexander T.\ Kristensson and Matthias Wilhelm}
\par\endgroup
\vspace{1em}
\begingroup\itshape
Niels Bohr International Academy, Niels Bohr Institute, Copenhagen University,\\
Blegdamsvej 17, 2100 Copenhagen \O{}, Denmark\\

\par\endgroup
\vspace{1em}
\begingroup\ttfamily
kristensson@nbi.ku.dk, matthias.wilhelm@nbi.ku.dk \\
\par\endgroup
\vspace{2.5em}
\endgroup

\begin{abstract}
\noindent
We study the thermodynamics of the maximally supersymmetric Yang-Mills theory with gauge group $\text{U}(N)$ on $\mathbb{R}\times S^3$, dual to type IIB superstring theory on AdS${}_5\times S^5$.
While both theories are well-known to exhibit Hagedorn behavior at infinite $N$, 
we find evidence that this is replaced by Lee-Yang behavior at large but finite $N$: the zeros of the partition function condense into two arcs in the complex temperature plane that pinch the real axis at the temperature of the confinement-deconfinement transition.
Concretely, we demonstrate this for the free theory via exact calculations of the (unrefined and refined) partition functions at $N\leq 7$ for the $\mathfrak{su}(2)$ sector containing two complex scalars, as well as at $N\leq 5$ for the $\mathfrak{su}(2|3)$ sector containing 3 complex scalars and 2 fermions. 
In order to obtain these explicit results, we use a Molien-Weyl formula for arbitrary field content, utilizing the equivalence of the partition function with what is known to mathematicians as the Poincar\'e series of trace algebras of generic matrices.
Via this Molien-Weyl formula, we also generate exact results for larger sectors.
\end{abstract}


\thispagestyle{empty}

\newpage
\hrule
\tableofcontents
\afterTocSpace
\hrule
\afterTocRuleSpace

\section{Introduction}
\label{sec: intro}

Gauge theories exhibit a rich thermodynamic structure, much of which is still to be understood.
This is even the case for what might be the simplest gauge theory, namely the maximally ($\mathcal{N}=4$) supersymmetric Yang-Mills (SYM) theory with gauge group $\text{U}(N)$, on $\mathbb{R}\times S^3$.
Via the AdS/CFT correspondence \cite{Maldacena:1997re}, a dual description of $\mathcal{N}=4$ SYM theory is given by type IIB superstring theory on AdS${}_5\times S^5$, which has been used to study the thermodynamic properties of both theories from early on \cite{Witten:1998zw}.

Gauß' law dictates that the states on a compact space such as $S^3$ are color singlets, leading to a phase transition in $\mathcal{N}=4$ SYM theory that bears resemblance to the confinement-deconfinement transition in QCD. 
The compact space $S^3$ allows also for a direct comparison between the conformal $\mathcal{N}=4$ SYM theory and confining theories, such as QCD \cite{Aharony:2003sx}; the finite radius $R_{S^3}$ acts as an effective infrared cutoff, limiting the running of couplings in the latter case, and making perturbation theory applicable in confining theories when tuned sufficiently small.
In conformal theories, the product of the temperature $T$ and $R_{S^3}$ yields a dimensionless quantity on which thermodynamic quantities can depend. In the following, we will thus set $R_{S^3}=1$ in the understanding that the dependence on $R_{S^3}$ can trivially be restored.
Via the  AdS/CFT correspondence, the confinement-deconfinement phase transition in $\mathcal{N}=4$ SYM theory is conjectured \cite{Witten:1998zw} to be dual to the Hawking-Page phase transition \cite{Hawking:1982dh} between a gas of gravitons (or closed strings) and a black hole.

A theoretical description of thermal physics is based on the partition function
\begin{equation}
\label{eq: unrefined partition function intro}
 \mathcal{Z}(T)=\sum_{\text{states}}e^{-E/T}\,,
\end{equation}
where the sum is over all states, $E$ denotes the energy of a given state and $T$ is the temperature in units of the Boltzmann constant. 
For example, the phase transition at temperature $T_c$ can be detected by looking at the scaling of the free energy with respect to $N$ \cite{Witten:1998zw}:
\begin{equation}
\label{eq: phase transition}
 F(T)=-T\log \mathcal{Z}(T)\sim \begin{cases}
                                 1 &\text{for }T<T_c\,,\\
                                 N^2 &\text{for }T>T_c\,.
                                \end{cases}
\end{equation}
A more detailed description is obtained by including also chemical potentials $\Omega_i$ for the two spins $S_1$, $S_2$ and three R-charges $J_1$, $J_2$, $J_3$ of $\mathcal{N}=4$ SYM theory. This yields the refined partition function 
\begin{equation}
\label{eq: refined partition function intro}
 \mathcal{Z}(T)=\sum_{\text{states}}e^{-(E -\sum_{i=1}^3 \Omega_i J_i - \sum_{a=1}^2 \Omega_{a+3} S_a)/T}\,.
\end{equation}

Using the state-operator map, the states on $\mathbb{R}\times S^3$ can be described by gauge-invariant local composite operators on flat Minkowski space $\mathbb{R}^{1,3}$; their energies $E$ are then given by the scaling dimensions $\Delta$. At tree level, the scaling dimension of an operator is simply the operator's engineering dimension, but quantum corrections shift it in the interacting theory.
Gauge-invariant local composite operators are built as traces of products of fields that transform covariantly under gauge transformations; moreover, products of such traces are again gauge invariant.
Operators containing one trace are conventionally called single-trace operators, whereas operators containing more than one trace are called multi-trace operators.

At infinite $N$, no relations exist between single- and multi-trace operators and a basis of the latter can be generated from the former, such that it suffices to consider single-trace operators. 
Due to their cyclicity, single-trace operators can be thought of as necklaces built from a set of beads, and counted using P\'olya theory. This observation was used by Sundborg \cite{Sundborg:1999ue} to calculate the free partition function of $\mathcal{N}=4$ SYM theory at infinite $N$.
This combinatorial approach was later extended to include the first correction in the 't Hooft coupling $\lambda=g_{\YM}^2N$, to chemical potentials and to related theories \cite{Spradlin:2004pp,Yamada:2006rx,Harmark:2006di,GomezReino:2005bq,Suzuki:2017ipd,Aharony:2005bq,Aharony:2006rf,Mussel:2009uw,Fokken:2014moa,Ramgoolam:2018epz}.

An important property displayed by the partition function at infinite $N$ is Hagedorn behavior -- an exponential growth of the density of states with energy.
This can even be seen without knowledge of the full partition function, by a rough estimate of the density of states.
Following Ref.\ \cite{Aharony:2003sx}, consider as a toy model the so-called $\mathfrak{su}(2)$ sector of $\mathcal{N}=4$ SYM theory, which is constructed from two complex scalars, say $X=\phi_1+i\phi_4$ and $Y=\phi_2+i\phi_5$, each having bare scaling dimension and thus energy $E=\Delta=1$. The number of single-trace states built from $E$ of these scalars, $\rho(E)$, can be estimated to be $\frac{2^E}{E}\leq \rho(E) \leq 2^E$, where the division by $E$ in the lower bound overaccounts for the fact that a trace is invariant under the $E$ cyclic permutations of the $E$ matrices in it, and this fact is neglected in the upper bound.
Both of the bounds, and thus also $\rho(E)$, grow exponentially with $E$ as $\rho(E)\sim e^{E\log2}$, such that a single-trace partition function $Z(T)=\sum_E \rho(E) e^{-E/T}$ given by this density of states diverges at the Hagedorn temperature $T_H^{\mathfrak{su}(2),\text{tree}}=1/\log2$, and so does the full multi-trace partition function.%
\footnote{The restriction to a subsector of the full theory can be thought of as the following limit in the partition function. Take $(\Omega_1,\dots,\Omega_5)=(n_1\Omega,\dots,n_5\Omega)$ for some $n_1,\dots n_5$ and consider $\Omega\to1$ with $T'=T/(1-\Omega)$ fixed. Then, only states from a specific subsector survive in the sum over all states that defines the partition function. To select for instance the $\mathfrak{su}(2)$ sector, one can choose $(n_1,\dots,n_5)=(1,1,0,0,0)$. This reasoning can also be extended to non-vanishing coupling to obtain a decoupling limit, see e.g.\ Ref.\ \cite{Harmark:2007px}.}
The Hagedorn behavior is present in all (non-trivial) subsectors of the theory, including the full theory; the different field content only effects the value of the Hagedorn temperature.
The Hagedorn temperature of the full $\mathcal{N}=4$ SYM theory was calculated via the partition function at tree level \cite{Sundborg:1999ue} and at first order in $\lambda$ \cite{Spradlin:2004pp}.
In the dual string theory, the Hagedorn behavior of the gauge theory is reflected in the well-known Hagedorn behavior of free (or tree-level) string theory \cite{Atick:1988}.

In the planar limit, the scaling dimensions of all operators in $\mathcal{N}=4$ SYM theory are in principle known via integrability; see Refs.\ \cite{Beisert:2010jr,Bombardelli:2016rwb} for reviews.%
\footnote{Interestingly, also the superconformal index could be calculated via a Bethe ansatz \cite{Benini:2018ywd}.}
Using integrability, also the Hagedorn temperature of $\mathcal{N}=4$ SYM theory and thus type IIB superstring theory on AdS${}_5\times S^5$ can be calculated at any value of the coupling \cite{Harmark:2017yrv,Harmark:2018red}; explicit results exist both numerically at finite coupling as well as analytically up to the seventh order in $\lambda$ \cite{talkIGST18,Harmark:2019} at weak coupling. At large $\lambda$, it asymptotes to the Hagedorn temperature of type IIB superstring theory in ten-dimensional Minkowski space \cite{Harmark:2018red} calculated in Ref.\ \cite{Sundborg:1984uk}.

At finite $N$, so-called trace relations exist that relate single-trace operators with more that $N$ fields to sums of multi-trace operators.\footnote{A basis of operators at finite $N$ is given by so-called Schur operators \cite{Corley:2001zk,Brown:2007xh,Bhattacharyya:2008rb}.} 
As a consequence, the thermodynamic behavior at large but finite $N$ drastically differs from the one at infinite $N$.
In particular, the trace relations cut off the exponential growth of the density of states with the energy for $E>N$, such that no Hagedorn behavior occurs for finite $N$, no matter how large.%
\footnote{This is also consistent with the known fact from statistical physics that the partition function of a system with a finite number of degrees of freedom on a compact space cannot have divergences at finite temperature.}
While the low-temperature phase ceases to exist at the Hagedorn temperature, the (confinement-deconfinement) phase transition at large but finite $N$ occurs at the lower critical temperature $T_c\leq T_H$.
On the dual string-theory side, finite $N$ corresponds to a non-vanishing string coupling, which allows for the Hawking-Page transition to a black hole.%
\footnote{Further aspects of the thermodynamic behavior at finite $N$, also for further systems, have been studied in Refs.\ \cite{Thorn:1980iv,Thorn:2015bia,Raha:2017jgv,Curtright:2017pfq,Raha:2019gyr,Hanada:2016pwv,Berenstein:2018lrm,Berenstein:2018hpl,Bergner:2019rca,Hanada:2019czd,Hanada:2019rzv,ArabiArdehali:2019orz,Hanada:2020uvt,Watanabe:2020ufk}.
}

At tree level, the partition function at finite $N$ can be written as a power series in $x=e^{-1/T}$, where the coefficients are written in terms of Littlewood-Richardson coefficients counting the number of color singlets in a tensor product of adjoint $\text{U}(N)$ representations, or characters of the symmetric group $S_n$ \cite{Aharony:2003sx,Dutta:2007ws}.
This sum representation, to which we refer as `character formula', was used to obtain closed expressions for the free unrefined partition function in the $\mathfrak{su}(2)$ sector for $N\leq5$ (as well as for larger numbers of complex scalars) \cite{Harmark:2014mpa} and extended to the refined partition function for $N\leq 4$ \cite{Vardinghus:2015}, starting from an ansatz as a rational function in $x$.

A powerful alternative mathematical formulation of the partition function is obtained by translating this problem from representation theory to invariant theory: in this context, the refined tree-level partition function in the $\mathfrak{su}(2)$ sector is the bigraded Poincar\'e series of $\text{U}(N)$ invariants of two Hermitian matrices, which is the same as the bigraded Poincar\'e series of $\text{GL}(N)$ invariants of two generic matrices.
Invariant theory provides an integral representation for this Poincar\'e series and its extension to more matrices, known as the Molien-Weyl formula, which proves to be an easier avenue to closed expressions of the partition functions at finite $N$ than the power-series representation.
In particular, the bigraded Poincar\'e series of $\text{GL}(N)$ invariants of two generic $N\times N$ matrices was calculated for $N\leq 6$ already long ago \cite{FORMANEK1981105,teranishi1986,teranishi1987,dhokovic2007}.%
\footnote{In the context of partition functions in gauge theories, such a representation was used in Ref.\ \cite{Dolan:2007rq}.}

In this paper, we employ the Molien-Weyl formula to obtain the tree-level partition function in the $\mathfrak{su}(2)$ sector at $N=7$.
More importantly, we find strong evidence that the Hagedorn temperature at infinite $N$ is replaced by so-called Lee-Yang behavior \cite{Yang:1952be} at large but finite $N$: the zeros of the partition function appear to condense into two arcs in the complex $x=e^{-1/T}$ plane that pinch the real axis at the critical temperature $T_c$. In the case of the $\mathfrak{su}(2)$ sector at tree level, we find $T^{\mathfrak{su}(2),\text{tree}}_c= T^{\mathfrak{su}(2),\text{tree}}_H=1/\log2$, which is consistent with the findings of Ref.\ \cite{Aharony:2003sx} that $T_c$ and $T_H$ coincide in the free theory. 
This main result of our paper is also reminiscent of the work of Witten and Maloney \cite{Maloney:2007ud}, who found that the Hawking-Page phase transition in three-dimensional quantum gravity is of Lee-Yang type.
Moreover, Lee-Yang zeros are used in the context of lattice QCD to detect phase transitions, see e.g.\ Refs.\ \cite{Nagata:2012tc,Nakamura:2013ska,Nagata:2014fra,Wakayama:2018wkc,Wakayama:2019hgz}.
Finally, we use the generalization of the Molien-Weyl formula to a general field content \cite{Dolan:2007rq} to generate many further explicit results.

The remainder of this paper is structured as follows. In Section \ref{sec: partition functions}, we review the infinite sum representation of the free partition function at finite $N$ and recall how it reproduces the infinite $N$ results as $N\to\infty$. Moreover, we present the generalized Molien-Weyl formula for the finite-$N$ partition function, which is an integral representation.
In Section \ref{sec: su(2) sector and Lee-Yang behavior}, we demonstrate how the Molien-Weyl formula can be used to obtain explicit expressions for the free partition function in the $\mathfrak{su}(2)$ sector at fixed $N$, obtaining new results at $N=7$. Furthermore, we show how the zeros of these partition functions condense in two arcs in the complex $x=e^{-1/T}$ plane, indicating Lee-Yang behavior.
In Section \ref{sec: larger sectors}, we proceed to larger sectors. We obtain explicit results for the free partition function in the fermionic $\mathfrak{su}(2|3)$ sector for $N\leq5$, which confirm the Lee-Yang behavior observed in the purely bosonic $\mathfrak{su}(2)$ sector in the previous section. Moreover, we also give the Molien-Weyl formula in the non-compact bosonic $\mathfrak{sl}(2)$ sector, in the higher-rank, non-compact, fermionic $\mathfrak{psu}(1,1|2)$ sector, as well as in the full theory. As a proof of principle, we evaluate the integral formulas in the $\mathfrak{sl}(2)$ and $\mathfrak{psu}(1,1|2)$ sector explicitly for $N=2$.
Our conclusion and outlook can be found in Section \ref{sec: conclusion}.
Three appendices provide details on the derivation of the character formula (Appendix \ref{app:deriv_character_formula}), the generalized Molien-Weyl formula (Appendix \ref{app:deriv_Molien_Weyl_formula}) and the calculation of the partition function in the $\mathfrak{psu}(1,1|2)$ sector at $N=2$ (Appendix \ref{app:psu(112)_N2}).

\section{Partition functions at infinite and finite \texorpdfstring{$N$}{N}}
\label{sec: partition functions}

In this section, we present two different (but mathematically equivalent) methods to evaluate the (refined) partition function for free gauge theories on $\mathbb{R} \times S^3$. We take these gauge theories to contain a general number and type of fields transforming in the adjoint representation of the gauge group $\text{U}(N)$.

\subsection{The Character Formula}

The first method takes the form of an infinite sum and was developed in Refs.\ \cite{Aharony:2003sx,Dutta:2007ws}. Since it makes use of group characters, we refer to it as the `character formula'. We review its derivation in Appendix \ref{app:deriv_character_formula}, and simply quote the resulting expression here:
\begin{equation} \label{eq:generating_function}
	\mathcal{Z}(\beta) = \sum_{n=0}^\infty \sum_{k\vdash n} \sum_{r\vdash n} \prod_{j=1}^n \frac{z(j\beta)^{k_j}}{k_j! j^{k_j}} \abs{\chi_r(k)}^2 \ ,
\end{equation}
where $\beta=1/T$, $k$ and $r$ are both integer partitions of $n$ labeling the irreducible representations of the symmetric group $S_n$ and $\text{U}(N)$, respectively, and $\chi_r(k)$ is the character of a group element with cycle structure $k$
in representation $r$.%
\footnote{Recall that the characters of $S_n$ only depend on the conjugacy class of a group element, which is specified by the cycle structure \cite{Zee:2016fuk}.}
The Young tableau corresponding to the partition $r$ is limited to have at most $N$ rows.

The (refined) single-particle partition function $z(\beta)$ is obtained by summing over all fields of the theory:
\begin{equation}
 z(\beta)=\sum_{\text{fields}}e^{-\beta(\Delta -\sum_{i=1}^3 \Omega_i J_i - \sum_{a=1}^2 \Omega_{a+3} S_a)}\,.
\end{equation}
In the case of $\mathcal{N}=4$ SYM theory, these are the fields corresponding to the spins in the spin-chain picture, transforming in the so-called singleton representation of the symmetry algebra $\mathfrak{psu}(2,2|4)$; see e.g.\ Ref.\ \cite{Minahan:2010js}.
In the $\mathfrak{su}(2)$ sector, we only have two complex scalar fields $X=\phi_1+i\phi_4$ and $Y=\phi_2+i\phi_5$, which have vanishing spins and R charges $J_X=\delta_{X1}$, $J_Y=\delta_{Y2}$. The (refined) single-particle partition function in this sector is thus $z(\beta)=e^{-\beta(1-\Omega_1)}+e^{-\beta(1-\Omega_2)}$. 
If the theory contains fermions, $z(j\beta)$ in Eq.\ \eqref{eq:generating_function} is to be understood as shorthand: 
\begin{equation}
 z(j\beta) \equiv z_B(j\beta) - (-1)^j z_F(j\beta)\,,
\end{equation}
where $z_B(\beta)$ and $z_F(\beta)$ are the single-particle partition functions for the bosonic and fermionic fields, respectively:
\begin{equation}
 z_{B/F}(\beta)=\sum_{\text{bosonic/fermionic fields}}e^{-\beta(\Delta -\sum_{i=1}^3 \Omega_i J_i - \sum_{a=1}^2 \Omega_{a+3} S_a)}\,.
\end{equation}

The single-particle partition function starts at least at order $\mathcal{O}(x^1)$ in $x\equiv e^{-\beta}\equiv e^{-1/T}$, since the minimal bare scaling dimension of a field in four dimensions is $\Delta=1$. 
This means that we can evaluate the partition function up to order $\mathcal{O}(x^L)$ by only calculating the terms with $n\leq L$, allowing a method for calculation of the power expansion of the partition function. If the structure of the partition function is simple enough, it is possible to guess the full exact function from a limited number of terms in the power expansion.
This method was used in Ref.\ \cite{Harmark:2014mpa} for the $\mathfrak{su}(q)$ `sector'\footnote{%
$\mathcal{N}=4$ SYM theory contains of course only three complex scalars, and the $\mathfrak{su}(3)$ sector is not closed beyond one-loop, closing to the larger $\mathfrak{su}(2|3)$ sector which we treat in Section \ref{subsec: su(23)}.
} (with unrefined single-particle partition function $z(x)=q x$) to calculate exact, unrefined partition functions for $N\leq 5$ in the case of $q=2$ and for $N\leq 3$ in the cases $q=3,4,5$, based on an ansatz as a rational function in $x$.
The extension to refined partition functions with non-zero chemical potentials was investigated in Ref.\ \cite{Vardinghus:2015} for $N\leq 4$ in the case $q=2$, for $N\leq3$ in the case $q=3$ and for $N\leq2$ in the case $q=4$. 
The disadvantage of the character formula is that the exact expressions for the partition functions can be very complicated, especially as $N$ increases. In this case, the exact partition functions are very hard to guess from the power expansion as many more terms are needed; cf.\ our explicit results in Sections \ref{sec: su(2) sector and Lee-Yang behavior} and \ref{sec: larger sectors}.%
\footnote{Moreover, the computational cost of any given order also increases with $N$.}

Let us now consider what happens in the limit of $N$ going to infinity. If $n\leq N$, the sum over partitions $r$ is unrestricted and we can use the row orthogonality of the characters to show \cite{Zee:2016fuk}:
\begin{equation}
	\sum_{r\vdash n} \abs{\chi_r(k)}^2 = \prod_{j=1}^n k_j! j^{k_j} \,.
\end{equation}
For infinite $N$, we can insert this back into Eq.\ \eqref{eq:generating_function}, which yields the simple expression \cite{Sundborg:1999ue}
\begin{equation} \label{eq:partition_planar}
	\mathcal{Z}_{N\rightarrow\infty}(\beta) = \sum_{n=0}^\infty \sum_{k\vdash n}  \prod_{j=1}^n z(j\beta)^{k_j} \ .
\end{equation}
Notice that the sums over $n$ and partitions $k\vdash n$ can be replaced by an infinite set of sums over $k_j$ from $1$ to $\infty$. These $k_j$ correspond to the number of rows with length $j$ in the Young tableau corresponding to $k$. Then we have the simple expression
\begin{equation}\label{partition_inf_N}
	\mathcal{Z}_{N\rightarrow\infty}(\beta) = \prod_{j=1}^\infty \sum_{k_j=1}^\infty z(j\beta)^{k_j} = \prod_{j=1}^\infty \frac{1}{1-z(j\beta)} \, .
\end{equation}
The partition function in this limit clearly diverges when $z(j\beta)=1$ for $j=1,2,\dotsc$. The temperature of the lowest pole at $z(\beta_H=1/T_H)=1$ corresponds to the Hagedorn temperature for infinite $N$ \cite{Sundborg:1999ue}, which coincides with the  confinement-deconfinement temperature in the free theory \cite{Aharony:2003sx}.

\subsection{The Molien-Weyl Formula}

Another method to calculate partition functions for gauge theories is given by the Molien-Weyl formula.
In general, the Molien-Weyl formula is a way to generate the Hilbert or Poincar\'e series for a ring of group invariants.
In the context of high-energy physics, it has been applied for example to count chiral gauge-invariant operators in SQCD (see e.g.\ Ref.\ \cite{Gray:2008yu} and references therein) as well as for BSM-EFT (see e.g.\ Ref.\ \cite{Banerjee:2020bym} and references therein).

An explicit Molien-Weyl formula for the free partition function of $\mathcal{N}=4$ SYM theory with gauge group $\text{U}(N)$ and any field content was given in Ref.\ \cite{Dolan:2007rq}. We review its derivation in our conventions and notation in Appendix \ref{app:deriv_Molien_Weyl_formula}.
The resulting expression is%
\footnote{The partition function for gauge group $\text{SU}(N)$ can be trivially obtained by replacing $\qty(\mathcal{Z}_{N=1}(x))^N\to\qty(\mathcal{Z}_{N=1}(x))^{N-1}$ in Eq.\ \eqref{eq:Molien_Weyl_formula}.}
\begin{equation} \label{eq:Molien_Weyl_formula}
	\mathcal{Z}_N(x) = \qty(\mathcal{Z}_{N=1}(x))^N \frac{1}{(2\pi i)^{N-1}} \oint_{|t_1|=1} \frac{\dd t_1}{t_1} \cdots \oint_{|t_{N-1}|=1} \frac{\dd t_{N-1}}{t_{N-1}} 
	\prod_{1\leq k\leq r \leq N-1}  \frac{1-t_{k,r}}{\phi_{k,r}} \ ,
\end{equation}
where $x\equiv e^{-\beta}= e^{-\frac{1}{T}}$, $t_{k,r}=t_k t_{k+1}\cdots t_r$ and
\begin{align}
\label{eq: general U(1) parttition function}
	\mathcal{Z}_{N=1}(x) &= \frac{\prod_{\text{fermionic fields}} (1+x^{\tilde{\Delta}})}{\prod_{\text{bosonic fields}} (1-x^{\tilde{\Delta}})} \, ,
\\
	\phi_{k,r} &= \frac{\prod_{\text{bosonic fields}} (1-x^{\tilde{\Delta}}t_{k,r})(1-x^{\tilde{\Delta}}t_{k,r}^{-1})}{\prod_{\text{fermionic fields}} (1+x^{\tilde{\Delta}}t_{k,r})(1+x^{\tilde{\Delta}}t_{k,r}^{-1})} \, ,
\end{align}
with
\begin{equation}
 \tilde{\Delta}=\Delta -\sum_{i=1}^3 \Omega_i J_i - \sum_{a=1}^2 \Omega_{a+3} S_a\,.
\end{equation}
The products run over all bosonic respectively fermionic fields in the theory with $\Delta$ denoting the conformal dimensions of the fields, $S_a$ their spin and $J_i$ their R-charge.
Note that, compared to the character formula, the field content now appears as a product instead of a sum.
The integrals all run over the unit circle in the complex planes of the $t_j$. Because of this, we can use residue theory to replace each integral by a sum over all residues within the unit circle.

Note that the partition function in the case of $N=1$, $\mathcal{Z}_{N=1}(x)$ in Eq.\ \eqref{eq: general U(1) parttition function}, simply contains one Bose-Einstein factor for each bosonic field and one Fermi-Dirac factor for each fermionic field. 
Restricting furthermore to only bosons, the algebra of gauge invariants in this case is freely generated from $\tr(\Phi)$, where $\Phi$ is running over all fields of the theory; it is simply a polynomial ring.

As an important special case, consider again the $\mathfrak{su}(2)$ sector composed of two scalar fields $X=\phi_1+i\phi_4$ and $Y=\phi_2+i\phi_5$ with conformal dimensions $\Delta_X=\Delta_Y=1$, vanishing spins and R charges $J_X=\delta_{X1}$, $J_Y=\delta_{Y2}$. The Molien-Weyl formula \eqref{eq:Molien_Weyl_formula} then reduces to the form \cite{teranishi1986,teranishi1987,dhokovic2007}:
\begin{equation} \label{eq:Molien_Weyl_su(2)}
\begin{aligned}
	\mathcal{Z}_N^{\mathfrak{su}(2)}(x_1,x_2) &= \frac{1}{(1-x_1)^N(1-x_2)^N} \\& \phaneq \times\frac{1}{(2\pi i)^{N-1}} \oint_{|t_1|=1} \frac{\dd t_1}{t_1} \cdots \oint_{|t_{N-1}|=1} \frac{\dd t_{N-1}}{t_{N-1}}  \prod_{1\leq k\leq r\leq N-1} \frac{1-t_{k,r}}{\phi_{k,r}} \, ,
\end{aligned}
\end{equation}
where
\begin{equation}
	\phi_{k,r} = (1-x_1t_{k,r})(1-x_2t_{k,r})(1-x_1t_{k,r}^{-1})(1-x_2t_{k,r}^{-1}) \ .
\end{equation}
Moreover, we now used the parameters $x_1=e^{-(1-\Omega_1)\beta}$, $x_2=e^{-(1-\Omega_2)\beta}$ as shorthand notation for the corresponding Boltzmann factors with distinct chemical potentials.

As was shown in the context of invariant theory in Ref.\ \cite{teranishi1986}, the Molien-Weyl formula in the $\mathfrak{su}(2)$ sector has an interesting property under inversion of its arguments:%
\footnote{This property has an immediate generalization to the $\mathfrak{su}(q)$ sector.}
\begin{equation}
\label{eq: temperature inversion su(2)}
 \mathcal{Z}_N^{\mathfrak{su}(2)}(1/x_1,1/x_2)=(-1)^{N-1} (x_1x_2)^{N^2}\mathcal{Z}_N^{\mathfrak{su}(2)}(x_1,x_2)\,.
\end{equation}
Let us briefly sketch the proof \cite{teranishi1986} of this property, restricting ourselves to the case of vanishing chemical potentials for simplicity, $x_1=x_2=x$.
In this case, the integrand \eqref{eq:Molien_Weyl_su(2)} is easily seen to transform with a factor $x^{2N^2}$ when sending $x\to1/x$. It has poles at $t_{N-1}\in r_{N-1}(x)\cup r_{N-1}(1/x)$, where $r_n(x)=\{x,\frac{x}{t_1},\frac{x}{t_1t_2},\dots,\frac{x}{t_1t_2\dots t_{n-1}}\}$. Since $0\leq T <\infty$ in the physical case, we usually assume that $|x|<1$, such that the poles $r_{N-1}(x)$ are within the unit circle.
The contour integral over the unit circle in $t_{N-1}$ then results in the sum over the residues at the poles $r_{N-1}(x)$. Sending $x\to 1/x$ results in the sum over the residues at the poles $r_{N-1}(1/x)$, which are the poles outside of the unit circle. As the sum of the residues at all poles vanishes via Cauchy's integration theorem, the two previous sums differ by a global sign. Having integrated in $t_{N-1}$ to $t_{a+1}$, the integrand in $t_a$ can be shown to have poles at $r_a(x)\cup r_a(x^2) \cup \dots \cup r_a(x^{N+1-a})\cup r_a(1/x)\cup r_a(1/x^2) \cup \dots \cup r_a(1/x^{N+1-a})$. Sending $x\to 1/x$ makes the contour integral in $t_a$ pick up the residues at the poles $r_a(1/x)\cup r_a(1/x^2) \cup \dots \cup r_a(1/x^{N+1-a})$ instead of $r_a(x)\cup r_a(x^2) \cup \dots \cup r_a(x^{N+1-a})$, again resulting in a global sign. Combining the effects of all $N-1$ contour integrations, we thus arrive at a global sign $(-1)^{N-1}$, concluding the proof.
Note that it was crucial for the proof that no residues at $t_i=0$ or $t_i=\infty$ existed, since these positions do not change when sending $x\to1/x$. For a different field content, the integrand of the Molien-Weyl formula can in fact have poles at  $t_i=0$ and $t_i=\infty$, such that no analog of the property \eqref{eq: temperature inversion su(2)} exists; we will encounter concrete examples of this in Section \ref{sec: larger sectors}.
The physical interpretation of the property \eqref{eq:Molien_Weyl_su(2)} is as follows: after taking a Casimir energy $x^{N^2}$ for $X$ and $Y$ into account, the partition function is symmetric (up to a sign) under temperature inversion $T\to -T$, a symmetry that was studied in detail for many other systems in Refs.\ \cite{Basar:2014mha,McGady:2017rzv,McGady:2018rmo}.%
\footnote{Interestingly, the temperature reflection symmetry is absent in this sector for infinite $N$, cf.\ Eq.\ \eqref{partition_inf_N}.}

\section{The \texorpdfstring{$\mathfrak{su}(2)$}{su(2)} sector and Lee-Yang behavior}
\label{sec: su(2) sector and Lee-Yang behavior}

Using the Molien-Weyl formula \eqref{eq:Molien_Weyl_su(2)}, the free refined partition functions in the $\mathfrak{su}(2)$ sector built from two complex scalars $X=\phi_1+i\phi_4$ and $Y=\phi_2+i\phi_5$ can be calculated as a sum of residues.
Explicit results for $N\leq 6$ were originally calculated via its identification with the Poincar\'e series of $\text{GL}(N)$ invariants of two generic matrices \cite{FORMANEK1981105,teranishi1986,teranishi1987,dhokovic2007}.%
\footnote{For $N\leq 5$, the results for the unrefined partition function have also been independently rederived in Ref.\ \cite{Harmark:2014mpa} by making an ansatz and determining the coefficients via the character formula \eqref{eq:generating_function}.}
We have extended these known results by calculating the refined partition function for $N=7$, using a specialized \textsc{Mathematica} code with a computation time of approximately 3 days on a desktop machine. We attach the refined partition functions in the ancillary file \texttt{su2partitionfunctions.m}. 

Below, we give the simpler, unrefined partition function, which are obtained by setting the chemical potentials to zero, i.e.\ setting $x_1 = x_2=x\equiv e^{-1/T}$.
For $N=1$, the matrices are numbers and the partition function is simply obtained via the geometric series:
\begin{equation}
	\mathcal{Z}^{\mathfrak{su}(2)}_{N=1}(x) = \frac{1}{(1-x)^2} \,,
\end{equation}
as can be seen immediately from Eq.\ \eqref{eq:Molien_Weyl_su(2)}.

For $N=2$, we have \cite{FORMANEK1981105}
\begin{equation}
	\mathcal{Z}^{\mathfrak{su}(2)}_{N=2}(x) = \frac{1}{(1-x)^2(1-x^2)^3} \,. 
\end{equation}
This has again the form of a geometric series. Indeed, it was shown in Ref.\ \cite{FORMANEK1981105} that the algebra of gauge invariants in this case is generated freely by $\tr(X)$, $\tr(Y)$, $\tr(X^2)$, $\tr(XY)$ and $\tr(Y^2)$. 

For $N=3$, we have \cite{teranishi1986}
\begin{equation}
\label{eq: su2 N=3}
	\mathcal{Z}^{\mathfrak{su}(2)}_{N=3}(x) = \frac{1+x^6}{(1-x)^2(1-x^2)^3(1-x^3)^4(1-x^4)}= \frac{1-x^2+x^4}{(1-x)^2(1-x^2)^4(1-x^3)^4} \,.
\end{equation}
Note that the partition function for $N=3$ has not the form of a geometric series, indicating that the algebra of gauge invariants is not freely generated. 
The elements $\tr(X)$, $\tr(Y)$, $\tr(X^2)$, $\tr(XY)$, $\tr(Y^2)$, $\tr(X^3)$, $\tr(X^2Y)$, $\tr(XY^2)$, $\tr(Y^3)$ and $\tr(X^2Y^2)$ are algebraically independent; they do not generate the full algebra though, only a subalgebra $C$. The full algebra of gauge invariants is obtained as $C\oplus (C \tr(XYX^2Y^2))$, as reflected in the numerator and denominator of the first form in Eq.\ \eqref{eq: su2 N=3}, cf.\ Ref.\ \cite{teranishi1986}.%
\footnote{The form of the partition function can also be analyzed via the so-called plethystic logarithm, see e.g.\ Refs.\ \cite{Benvenuti:2006qr,Feng:2007ur,Kimura:2009ur,Harmark:2014mpa}.
}

For $N=4$, we have \cite{teranishi1987}
\begin{equation}
	\mathcal{Z}^{\mathfrak{su}(2)}_{N=4}(x) = \frac{1-x-x^2+2x^4+2x^5-4x^7+2x^9+2x^{10}-x^{12}-x^{13}+x^{14}}{(1-x)^3(1-x^2)^4(1-x^3)^5(1-x^4)^5} \,,
\end{equation}
In this case, an identification of the minimal set of generating traces similar to $N=3$ becomes quickly quite tedious \cite{teranishi1987}.

For $N=5$ and $N=6$, we have \cite{dhokovic2007}
\begin{equation}
	\mathcal{Z}^{\mathfrak{su}(2)}_{N=5}(x) = \frac{P_{40}(x)}{(1-x)^{0}(1-x^2)^{6}(1-x^3)^{8}(1-x^4)^{6}(1-x^5)^{6}} \, ,
\end{equation}
with 
\begin{equation}
\begin{split}
	P_{40}(x)=\ &1+2x-6x^{3}-9x^{4}+2x^{5}+25x^{6}+38x^{7}+17x^{8}-34x^{9}\\&
	-68x^{10}-34x^{11}+73x^{12}+176x^{13}+171x^{14}+34x^{15}-127x^{16}\\&
	-156x^{17}-2x^{18}+218x^{19}+322x^{20}+218x^{21}-2x^{22}-156x^{23}\\&
	-127x^{24}+34x^{25}+171x^{26}+176x^{27}+73x^{28}-34x^{29}-68x^{30}\\&
	-34x^{31}+17x^{32}+38x^{33}+25x^{34}+2x^{35}-9x^{36}-6x^{37}+2x^{39}+x^{40} \, ,
\end{split}
\end{equation}
and
\begin{equation}
	\mathcal{Z}^{\mathfrak{su}(2)}_{N=6}(x) = \frac{P_{70}(x)}{(1-x)^5(1-x^2)^3(1-x^3)^6(1-x^4)^9(1-x^5)^7(1-x^6)^7} \ ,
\end{equation}
with
\begin{equation}
\begin{split}
	P_{70}(x)=\ &1-3x+3x^{2}-3x^{3}+3x^{4}+4x^{5}-2x^{6}-8x^{8}-8x^{9}\\&
	+11x^{10}+x^{11}+56x^{12}-24x^{13}+48x^{14}-69x^{15}-9x^{16}+2x^{17}\\&
	+78x^{18}+118x^{19}+223x^{20}+23x^{21}+158x^{22}-182x^{23}+221x^{24}\\&
	-42x^{25}+600x^{26}+365x^{27}+633x^{28}+324x^{29}+303x^{30}-31x^{31}\\&
	+484x^{32}+178x^{33}+1055x^{34}+518x^{35}+1055x^{36}+178x^{37}+484x^{38}\\&
	-31x^{39}+303x^{40}+324x^{41}+633x^{42}+365x^{43}+600x^{44}-42x^{45}\\&
	+221x^{46}-182x^{47}+158x^{48}+23x^{49}+223x^{50}+118x^{51}+78x^{52}\\&
	+2x^{53}-9x^{54}-69x^{55}+48x^{56}-24x^{57}+56x^{58}+x^{59}+11x^{60}\\&
	-8x^{61}-8x^{62}-2x^{64}+4x^{65}+3x^{66}-3x^{67}+3x^{68}-3x^{69}+x^{70} \, .
\end{split}
\end{equation}

For $N=7$, we found the following new result:
\begin{equation}
	\mathcal{Z}^{\mathfrak{su}(2)}_{N=7}(x) = \frac{P_{136}(x)}{(1-x)^0(1-x^2)^4(1-x^3)^8(1-x^4)^{12}(1-x^5)^{10}(1-x^6)^8(1-x^7)^8} \,,
\end{equation}
where
\begingroup
\allowdisplaybreaks
\begin{align}
	P_{136}(x)=\ &1+2 x+2 x^2-2 x^3-12 x^4-20 x^5-10 x^6+38 x^7+124 x^8+202 x^9+186 x^{10}\nonumber\\
	&-2 x^{11}-312 x^{12}-494 x^{13}-82 x^{14}+1364 x^{15}+3935 x^{16}+7080 x^{17}\nonumber\\
	&+9761x^{18}+11190 x^{19}+12188 x^{20}+16284 x^{21}+29980 x^{22}+61276 x^{23}\nonumber\\
	&+117046 x^{24}+200524 x^{25}+311834 x^{26}+452462 x^{27}+634771 x^{28}\nonumber\\
	&+891852 x^{29}+1284256 x^{30}+1896942 x^{31}+2828447 x^{32}+4174570 x^{33}\nonumber\\
	&+6021068 x^{34}+8452156 x^{35}+11582747 x^{36}+15602230 x^{37}+20815499 x^{38}\nonumber\\
	&+27651402x^{39}+36633392 x^{40}+48308938 x^{41}+63176460 x^{42}+81638768 x^{43}\nonumber\\
	&+104026405 x^{44}+130676134 x^{45}+162046094 x^{46}+198782434 x^{47}\nonumber\\
	&+241699563 x^{48}+291628632 x^{49}+349196244 x^{50}+414593302 x^{51}\nonumber\\
	&+487467350 x^{52}+566967546 x^{53}+651962894 x^{54}+741302716 x^{55}\nonumber\\
	&+834019828 x^{56}+929323032x^{57}+1026404662 x^{58}+1124098904 x^{59}\nonumber\\
	&+1220612186 x^{60}+1313438250 x^{61}+1399593383 x^{62}+1476059720 x^{63}\nonumber\\
	&+1540326729 x^{64}+1590748212 x^{65}+1626630377x^{66}+1647969244 x^{67}\nonumber\\
	&+1655036460 x^{68}+1647969244 x^{69}+1626630377 x^{70}+1590748212 x^{71}\nonumber\\
	&+1540326729 x^{72}+1476059720 x^{73}+1399593383 x^{74}+1313438250x^{75}\nonumber\\
	&+1220612186 x^{76}+1124098904 x^{77}+1026404662 x^{78}+929323032 x^{79}\nonumber\\
	&+834019828 x^{80}+741302716 x^{81}+651962894 x^{82}+566967546 x^{83}\nonumber\\
	&+487467350x^{84}+414593302 x^{85}+349196244 x^{86}+291628632 x^{87}\nonumber\\
	&+241699563 x^{88}+198782434 x^{89}+162046094 x^{90}+130676134 x^{91}\nonumber\\
	&+104026405 x^{92}+81638768x^{93}+63176460 x^{94}+48308938 x^{95}+36633392 x^{96}\nonumber\\
	&+27651402 x^{97}+20815499 x^{98}+15602230 x^{99}+11582747 x^{100}+8452156 x^{101}\nonumber\\
	&+6021068 x^{102}+4174570x^{103}+2828447 x^{104}+1896942 x^{105}+1284256 x^{106}\nonumber\\
	&+891852 x^{107}+634771 x^{108}+452462 x^{109}+311834 x^{110}+200524 x^{111}\nonumber\\
	&+117046 x^{112}+61276x^{113}+29980 x^{114}+16284 x^{115}+12188 x^{116}\nonumber\\
	&+11190 x^{117}+9761 x^{118}+7080 x^{119}+3935 x^{120}+1364 x^{121}-82 x^{122}\nonumber\\
	&-494 x^{123}-312 x^{124}-2x^{125}+186 x^{126}+202 x^{127}+124 x^{128}+38 x^{129}\nonumber\\
	&-10 x^{130}-20 x^{131}-12 x^{132}-2 x^{133}+2 x^{134}+2 x^{135}+x^{136} \, .
\end{align}
\endgroup

The numerators of all partition functions are given by palindromic polynomials in $x$.
This is a consequence of the partition functions being invariant under $x\rightarrow x^{-1}$ up to an overall factor of $(-1)^{N-1}x^{2N^2}$, as shown in Ref.\ \cite{teranishi1986}%
\footnote{The fact that the numerator polynomials are palindromic was also independently observed in examples in Ref.\ \cite{Harmark:2014mpa}.} and discussed below  Eq.\ \eqref{eq:Molien_Weyl_su(2)}.
As also mentioned below Eq.\ \eqref{eq:Molien_Weyl_su(2)}, this property can be interpreted as a symmetry under temperature reflection ($T\to-T$) up to a sign when including a Casimir energy of $N^2$, i.e.\ a Casimir energy of $1/2$ per real degree of freedom in each of the matrices $X$ and $Y$. 
The maximal orders of the (maximally canceled) polynomials increase rapidly with $N$: $0,0,4,14,40,70,136,\dots$ for $N=1,2,3,4,5,6,7,\dots$. The general structure of these polynomials for any $N$ remains to be found; as can be seen from Eq.\ \eqref{eq: su2 N=3}, the pattern is also prone to being obscured by cancellations between numerator and denominator.

The denominators of the partition functions are built from products of $(1-x^i)^{p_i}$, with $i=1,...,N$. The powers, $p_i$, all sum up to $N^2+1$. Thus for $x\rightarrow 1$ or $T\rightarrow \infty$, the partition functions all have limits of the form
\begin{equation}
	\mathcal{Z}_{N}(x) \approx \frac{a_{N}}{(1-x)^{N^2+1}} \qq{for} T\rightarrow \infty \ ,
\end{equation}
with $a_N$ some constant.
This property was shown in Ref.\ \cite{Harmark:2014mpa} in the more general setting of the $\mathfrak{su}(q)$ sector using Eq.\ \eqref{eq:generating_function}, and it was argued that at high temperatures the $\mathfrak{su}(q)$ sector on a compact space behaves as $(q-1)N^2+1$ independent harmonic oscillators.

\begin{figure}[tp]
\begin{center}
	\begin{tikzpicture}
	\begin{axis}[axis lines = middle,axis equal image,width=\linewidth]
		\addplot [only marks,mark=o,green] table {data/data_zeros4.txt};
		\addplot [only marks,mark=square,blue] table {data/data_zeros5.txt};
		\addplot [only marks,mark=triangle,red] table {data/data_zeros6.txt};
		\addplot [only marks,mark=diamond,black] table {data/data_zeros7.txt};
		\legend{$N=4$,$N=5$,$N=6$,$N=7$};
		\draw (axis cs:0,0) circle [radius=1];
	\end{axis}
	
	\end{tikzpicture}
	\caption{Zeros of the $\mathfrak{su}(2)$ partition functions for $N=4\dotsc 7$ plotted in the complex plane of $x=e^{-1/T}$. The right-most zeros appear to `condense' in two arcs pinching the real axis just around the Hagedorn temperature $x_H^{\mathfrak{su}(2),\text{tree}}=1/2$ of the infinite-$N$ theory. This hints at a phase transition of Lee-Yang type.}
	\label{fig:su(2)_zeros}
\end{center}
\end{figure}
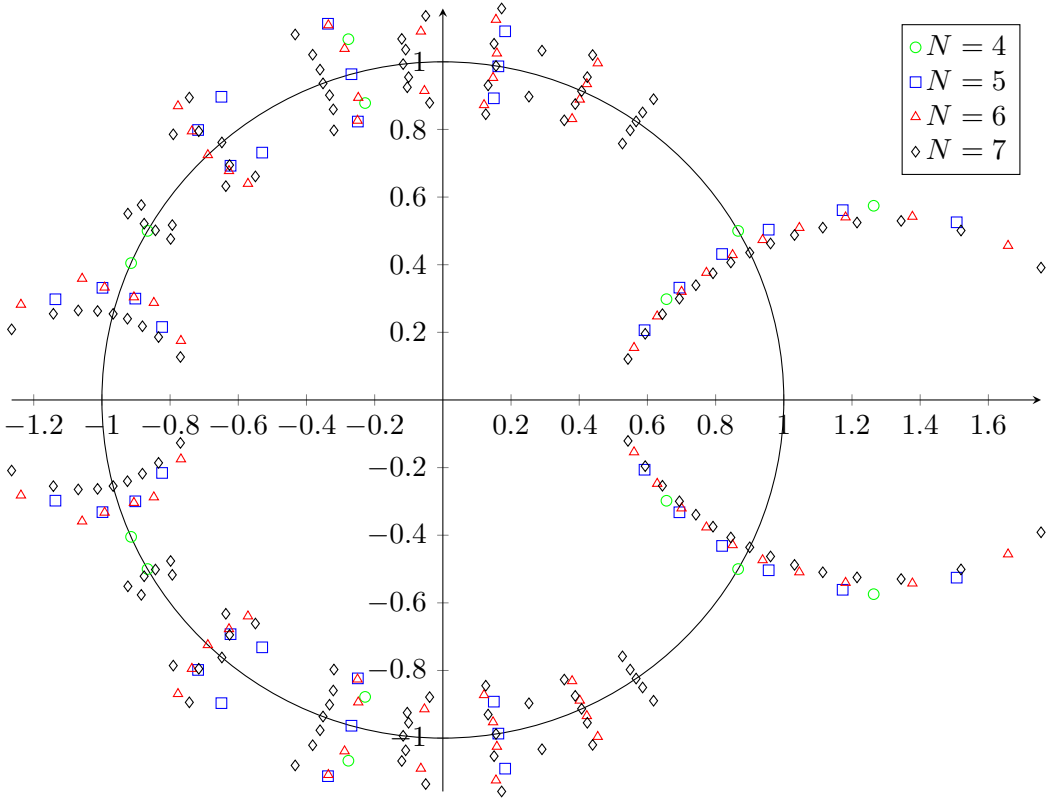

In Figure \ref{fig:su(2)_zeros}, we plot all zeros of the unrefined partition functions in the complex plane of $x=e^{-1/T}$ for $N=4,\dotsc,7$. Remarkably, the right-most group of zeros appears to `condense' in two arcs pinching the real axis, signaling a phase transition of Lee-Yang type \cite{Yang:1952be}.
The number of zeros on each arch for $N=1,\dotsc,7$ is $0,0,1,3,6,10,15$, which exactly follows the prescription $(N-1)(N-2)/2$. Along with the observation that the zero closest to the real axis appears to inch closer and closer for each value of $N$, this further solidifies the hypothesis that the zeros are condensing on these two arcs pinching the real axis.
Furthermore, the arcs appear to pinch the real axis just around the point $x^{\mathfrak{su}(2),\text{tree}}_c=1/2$ or $T_c^{\mathfrak{su}(2),\text{tree}}=1/\ln 2$, which is the Hagedorn temperature at infinite $N$.%
\footnote{Note that there appear to be also further, smaller arcs, away from the positive real axis. The most prominent of these appear to pinch the negative real axis at $x=-1/\sqrt{2}$, which is a solution to $z(j\beta)=1$ for $j=2$, i.e.\ it is the position of a higher Hagedorn pole, cf.\ Eq.\ \eqref{partition_inf_N} and the discussion below it. It would be interesting to develop a better understanding of these further arcs.}
This hints at the remarkable result that for large but finite $N$, the Hagedorn behavior is replaced by Lee-Yang behavior.
(Recall that the Hagedorn temperature is expected to coincide with the confinement-deconfinement temperature in the free theory, while the two are expected to differ at higher loop order \cite{Aharony:2003sx}.)
Via the AdS/CFT correspondence, also the Hawking-Page transition of type IIB superstring theory on AdS${}_5\times S^5$ should thus be of Yang-Lee type.

Our finding is reminiscent of the work of Maloney and Witten \cite{Maloney:2007ud}, who investigated the Hawking-Page phase transition in three-dimensional quantum gravity and found it to be of Lee-Yang type. 
With gravity being considerably simpler in three dimensions than in higher dimensions, it was also possible to rigorously prove the condensation of the zeros in this case.
Moreover, the condensation of Lee-Yang zeros is in fact a well-known phenomenon in the field of lattice QCD; see for instance Refs.\ \cite{Nagata:2012tc,Nakamura:2013ska,Nagata:2014fra,Wakayama:2018wkc,Wakayama:2019hgz}.

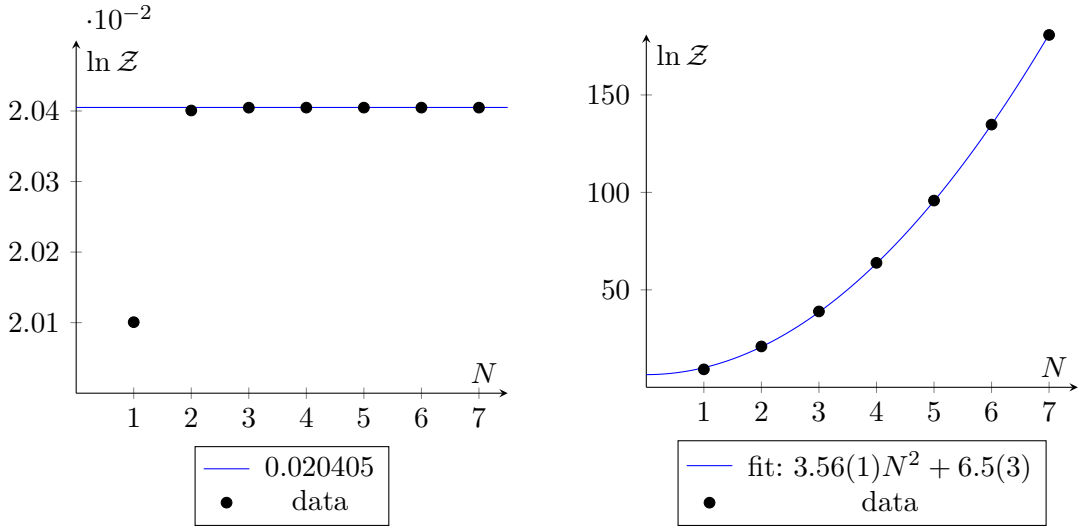
\begin{figure}[tp]
\begin{center}
	\begin{tikzpicture}
	\begin{axis}[xlabel=$N$,ylabel=$\ln \mathcal{Z}$,legend style={at={(0.5,-0.15)},anchor=north},axis lines = middle,xtick={0,1,2,3,4,5,6,7},xmin=0,xmax=7.5,ymin=0.02,ymax=0.0205,ytick={0.0200,0.0201,0.0202,0.0203,0.0204},
	width=0.48\linewidth]
		\addplot [domain=0:7.5, samples=100, color=blue] {0.020405};
		\addlegendentry{\ $0.020405$};
		\addplot [only marks] coordinates {(1,0.02010067209)(2,0.02040068666)(3,0.02040469602)(4, 0.02040474599)(5, 0.02040474795)(6, 0.02040474893)(7, 0.02040474697)};
		\addlegendentry{\ data};		
	\end{axis}
	\end{tikzpicture}
	\qquad
	\begin{tikzpicture}
	\begin{axis}[xlabel=$N$,ylabel=$\ln \mathcal{Z}$,legend style={at={(0.5,-0.15)},anchor=north},axis lines = middle,xtick={0,1,2,3,4,5,6,7},xmin=0,xmax=7.5,ymin=0,ymax=181,width=0.48\linewidth]
		\addplot [domain=0:7.5, samples=100, color=blue] {3.563*x^2 + 6.501};
		\addlegendentry{\ fit: $3.56(1) N^2 + 6.5(3)$}
    	\addplot [only marks] coordinates {(1,9.210340372)(2,20.96144701)(3,38.92508411)(4, 63.85946849)(5, 95.80415674)(6, 134.7745531)(7, 180.7791359)};
		\addlegendentry{\ data};	    	
	\end{axis}
	\end{tikzpicture}
	\caption{Plots of the logarithm of the $\mathfrak{su}(2)$ partition functions for $N=1,\dotsc,7$ in the low temperature regime (left) with $x=0.01$ and the high temperature regime (right) with $x=0.99$. The plot on the left includes a constant line at $\ln \mathcal{Z}^{\mathfrak{su}(2)}_{N=7}(x=0.01) \sim 0.020405$, for comparison. On the right, we fit the data to the function $c_2\cdot N^2 + c_0$, with an $R^2$-value of $0.999951$. As evident from the respective plots, the free energy scales as $N^0$/$N^2$ in the low/high temperature regime.}
	\label{fig:su(2)_free_energy_N}
\end{center}
\end{figure}

From the explicit partition functions for $N\leq 7$, we can also investigate the low- and high-temperature 
regimes of the $\mathfrak{su}(2)$ sector. In Figure \ref{fig:su(2)_free_energy_N}, we plot the logarithm of the partition functions for $N=1,\dotsc,7$, which is proportional to the free energy; cf.\ Eq.\ \eqref{eq: phase transition}. On the left, we have set $x=0.01$ corresponding to $T\approx 0.217$. On the right, we have $x=0.99$ corresponding to $T\approx 99.5$.
Using a fit, we observe that the free energy scales as $N^0$/$N^2$ in the low/high temperature regime, a clear sign of a confinement-deconfinement phase transition as described in Ref.\ \cite{Witten:1998zw}.

\section{Generalization to larger sectors}
\label{sec: larger sectors}

We now proceed to larger sectors.
Including fermions on top of scalars, we find further evidence for Lee-Yang behavior in the $\mathfrak{su}(2|3)$ sector in Subsection \ref{subsec: su(23)}.
We next go to the non-compact $\mathfrak{sl}(2)$ sector, which includes arbitrary numbers of covariant derivatives on a single complex scalar. While the Molien-Weyl formula now requires to sum over infinitely many residues, it can still be evaluated, and we demonstrate how this can be done explicitly for $N=2$ in Subsection \ref{subsec: sl(2)}.
In Subsection \ref{subsec: psu(112)}, we show that the same procedure also works for the non-compact higher-rank $\mathfrak{psu}(1,1|2)$ sector, which contains complex scalars, fermions and covariant derivatives.
Finally, we give the Molien-Weyl formula for the full theory in Subsection \ref{subsec: psu(224)}.

\subsection{\texorpdfstring{$\mathfrak{su}(2|3)$}{su(2|3)} sector}
\label{subsec: su(23)}

We now include fermions in the discussion, and investigate the $\mathfrak{su}(2|3)$ sector built from three complex scalars $X=\phi_1+i\phi_4$, $Y=\phi_2+i\phi_5$, $Z=\phi_3+i\phi_6$ and two fermions $\psi^{\alpha=1}_{4}$ and $\psi^{\alpha=2}_{4}$. 
While the scalars have classical scaling dimension $\Delta=1$, the fermions have classical scaling dimension $\Delta=3/2$, resulting in the single-particle partition functions
\begin{equation}
 z_B(x)=3x\,,\qquad z_F(x)=2x^{3/2}\,.
\end{equation}

Using Eq.\ \eqref{eq:Molien_Weyl_formula} in the $\mathfrak{su}(2|3)$ sector, we obtain the Molien-Weyl formula
\begin{equation} \label{eq:Molien_Weyl_su(2I3)}
\begin{split}
	\mathcal{Z}_N^{\mathfrak{su}(2|3)}(x) &=\frac{(1+x^{3/2})^{2N}}{(1-x)^{3N}} 
	 \frac{1}{(2\pi i)^{N-1}}  \oint_{|t_1|=1} \frac{\dd t_1}{t_1} \cdots \oint_{|t_{N-1}|=1} \frac{\dd t_{N-1}}{t_{N-1}}  \prod_{1\leq k\leq r\leq N-1} \frac{1-t_{k,r}}{\phi_{k,r}} \ ,
\end{split}
\end{equation}
where $x=e^{-1/T}$, $t_{k,r}=t_k t_{k+1}\cdots t_r$ and
\begin{equation}
	\phi_{k,r} = \frac{(1-xt_{k,r})^3(1-xt_{k,r}^{-1})^3}{(1+x^{3/2}t_{k,r})^2(1+x^{3/2}t_{k,r}^{-1})^2} \ .
\end{equation}
Note that we are restricting ourselves to the unrefined partition function here for simplicity; the refined one can be obtained in a similar way. 

As was the case with the $\mathfrak{su}(2)$ sector, we can evaluate this formula for low values of $N$ by replacing the integrals over the complex unit circle of $t_i$ with sums over the residues contained within it. 
Using this procedure, we were able to compute the $\mathfrak{su}(2|3)$ partition functions for $N \leq 5$ via \textsc{Mathematica}, as shown below. Since the fermions contribute with half-integer scaling dimensions, it is convenient to introduce $a\equiv \sqrt{x}=e^{-\frac{1}{2T}}$.
We include our full results in the ancillary file \texttt{su2slash3partitionfunctions.m}.

For $N=1$, we have 
\begin{equation}
	\mathcal{Z}_{N=1}^{\mathfrak{su}(2|3)}(a) = \frac{(1+a^3)^2}{(1-a^2)^3} \,.
\end{equation}
For $N=2$, we find 
\begin{equation}
	\mathcal{Z}_{N=2}^{\mathfrak{su}(2|3)}(a) = \frac{(1+a)^3(1+a^3)^4}{(1-a^2)^4(1-a^4)^5} \times P_{11}(a) \, ,
\end{equation}
where 
\begin{equation}
\begin{split}
	P_{11}(a) =1-3a+5{a}^{2}-9{a}^{3}+16{a}^{4}-18{a}^{5}+19{a}^{6}
	-21{
a}^{7}+17{a}^{8}-9{a}^{9}+7{a}^{10}-3{a}^{11}
 \,.
\end{split}
\end{equation}
For $N=3$, we find 
\begin{equation}
	\mathcal{Z}_{N=3}^{\mathfrak{su}(2|3)}(a) = \frac{(1+a)^{10}(1+a^3)^6}{(1-a^2)^4(1-a^4)^8(1-a^6)^7} \times P_{50}(a) \,,
\end{equation}
where
\begin{equation}
\begin{split}
	P_{50}(a) =\ &1-10a+54{a}^{2}-214{a}^{3}+698{a}^{4}-1972{a}^{5}+4977{a}^{6}-11458{a}^{7}+24401{a}^{8}\\
	&-48556{a}^{9}+90987{a}^{10}-161476{a}^{11}+272627{a}^{12}-439456{a}^{13}+678225{a}^{14}\\
	&-1004446{a}^{15}+1430189{a}^{16}-1960780{a}^{17}+2591626{a}^{18}-3305714{a}^{19}\\
	&+4072484{a}^{20}-4848720{a}^{21}+5581917{a}^{22}-6215426{a}^{23}+6695473{a}^{24}\\
	&-6978320{a}^{25}+7036587{a}^{26}-6863412{a}^{27}+6473870{a}^{28}-5902598{a}^{29}\\
	&+5199179{a}^{30}-4421048{a}^{31}+3625988{a}^{32}-2865264{a}^{33}+2178627{a}^{34}\\
	&-1591448{a}^{35}+1114790{a}^{36}-747174{a}^{37}+477876{a}^{38}-290730{a}^{39}\\
	&+167604{a}^{40}-91122{a}^{41}+46455{a}^{42}-22050{a}^{43}+9654{a}^{44}\\
	&-3852{a}^{45}+1380{a}^{46}-432{a}^{47}+114{a}^{48}-24{a}^{49}+3{a}^{50}
 \,.
\end{split}
\end{equation}
For $N=4$, we find
\begin{equation}
	\mathcal{Z}_{N=4}^{\mathfrak{su}(2|3)}(a) = \frac{(1+a)^{18}(1+a^3)^{11}}{(1-a^2)^6(1-a^4)^8(1-a^6)^{10}(1-a^8)^9} \times P_{119}(a) \ ,
\end{equation}
where $P_{119}(a)$ is a polynomial of degree $119$. Since $P_{119}(a)$ is one page long, we refrain from printing it here; the interested reader can find it in machine-readable form in the ancillary file \texttt{su2slash3partitionfunctions.m}.
For $N=5$, we find
\begin{equation}
	\mathcal{Z}_{N=5}^{\mathfrak{su}(2|3)}(a) = \frac{(1+a)^{18}(1+a^3)^{18}(1+a^5)^{10}}{(1-a^4)^{12}(1-a^6)^{16}(1-a^8)^{12}(1-a^{10})^{11}} \times P_{219}(a) \ ,
\end{equation}
where $P_{219}(a)$ is a two-and-a-half pages long polynomial of degree $219$ which can equally be found in the ancillary file \texttt{su2slash3partitionfunctions.m}. 

Note that the polynomials in the numerator are not palindromic, in contrast to the $\mathfrak{su}(2)$ sector.
Recall that in the $\mathfrak{su}(2)$ sector, the palindromicness is a consequence of the uniform transformation behavior of the partition function under $x\to x^{-1}$, which can be proven via the Molien-Weyl formula. Its proof required the poles of the integrand at zero and infinity (or rather the corresponding residues) to vanish. Indeed, the Molien-Weyl formula \eqref{eq:Molien_Weyl_su(2I3)} in the  $\mathfrak{su}(2|3)$ sector has non-vanishing residues at infinity, which spoil this property.%
\footnote{Note that temperature inversion symmetry is in general only expected to hold in the full theory \cite{McGady:2017rzv}.}

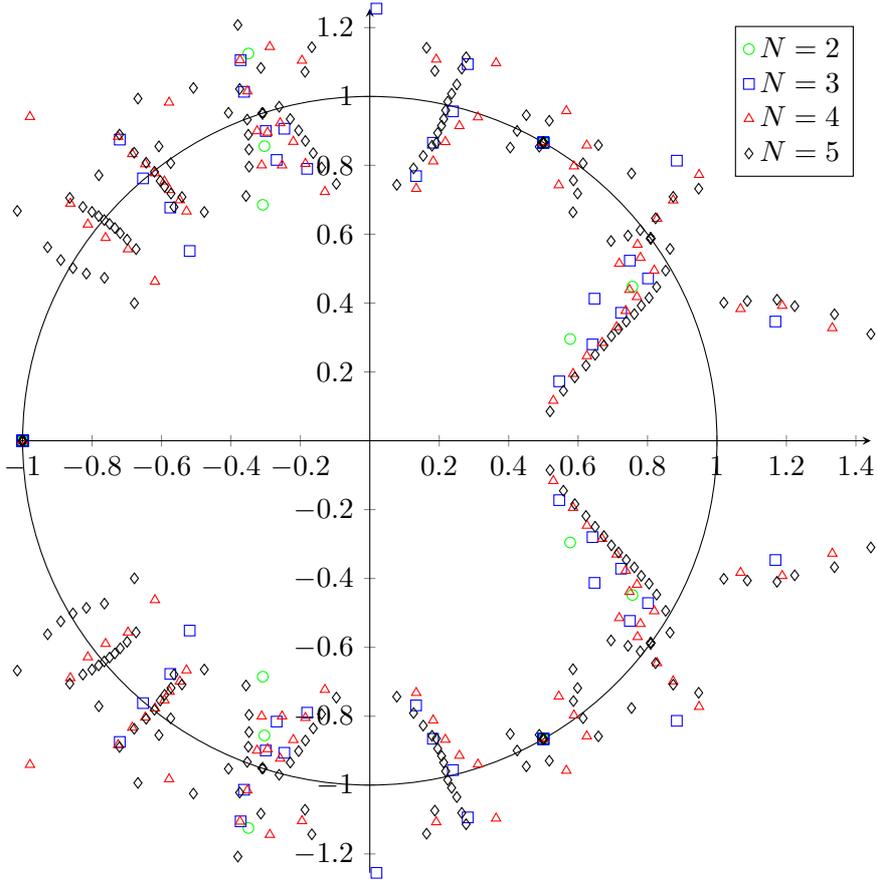
\begin{figure}[t]
\begin{center}
	\begin{tikzpicture}
	\begin{axis}[axis lines = middle,axis equal image,restrict x to domain=-1.3:1.5,restrict y to domain=-1.3:1.3,width=\linewidth]
		\addplot [only marks,mark=o,green] table {data/zeros_su2I3_2.txt};
		\addplot [only marks,mark=square,blue] table {data/zeros_su2I3_3.txt};
		\addplot [only marks,mark=triangle,red] table {data/zeros_su2I3_4.txt};
		\addplot [only marks,mark=diamond,black] table {data/zeros_su2I3_5.txt};
		\legend{$N=2$,$N=3$,$N=4$,$N=5$};
		\draw (axis cs:0,0) circle [radius=1];
	\end{axis}
	
	\end{tikzpicture}
	\caption{Zeros of the free partition functions in the $\mathfrak{su}(2|3)$ sector for $N=2\dotsc 5$ plotted in the complex plane of $a=e^{-\frac{1}{2T}}$. The right-most zeros appear to `condense' in two arcs pinching the real axis just around the Hagedorn temperature $a_H^{\mathfrak{su}(2|3),\text{tree}}=1/2$ of the infinite-$N$ theory. This hints at a phase transition of Lee-Yang type.}
	\label{fig:su(2I3)_zeros}
\end{center}
\end{figure}

In the $\mathfrak{su}(2)$ sector, we found compelling evidence that the Hagedorn behavior at infinite $N$ is replaced by Lee-Yang behavior at large but finite $N$, from observing a condensation of the zeros of the partition function in the complex $x=e^{-1/T}$ plane. In Figure \ref{fig:su(2I3)_zeros}, we plot the zeros of the $\mathfrak{su}(2|3)$ partition functions for $N\leq 5$ in the complex $a=e^{-\frac{1}{2T}}$ plane and observe a similar condensation of the zeros into two main arcs that appear to pinch the real axis.
From Eq.\ \eqref{partition_inf_N}, we see that the Hagedorn temperature of the $\mathfrak{su}(2|3)$ sector at infinite $N$ is given by the lowest positive solution to the equation $1=z(a)=z_B(a)+z_F(a)=3a^2+2a^3$, which is at $a_H^{\mathfrak{su}(2|3),\text{tree}}=1/2$ corresponding to $T_H^{\mathfrak{su}(2|3),\text{tree}} = 1/2\ln 2$. 
(Recall again that $T_c=T_H$ in the free theory \cite{Aharony:2003sx}.)
In Figure \ref{fig:su(2I3)_zeros}, the zeros appear to pinch the real axis exactly at $a_H^{\mathfrak{su}(2|3),\text{tree}}=1/2$, further supporting the statement that Lee-Yang behavior replaces Hagedorn behavior at finite $N$.

\subsection{\texorpdfstring{$\mathfrak{sl}(2)$}{sl(2)} sector}
\label{subsec: sl(2)}

The $\mathfrak{sl}(2)$ sector is built from one complex scalar, say $X=\phi_1+i\phi_4$, and a single light-like covariant derivative that can act on it, say $\mathcal{D}\equiv \mathcal{D}^{\mu}\sigma_{\mu}^{\alpha=1,\dot\alpha=1}$, resulting in fields of type $\mathcal{D}^{i-1}X$ for $i=1,2,3\dots$.\footnote{Recall that $\sigma_{\mu}=(\idm,\sigma_1,\sigma_2,\sigma_3)$, where $\sigma_i$ are the usual Pauli matrices.}
Since $\Delta_{\mathcal{D}^{i-1}X}=i$, we have the single-particle partition function
\begin{equation}
	z(x)= \sum_{i=1}^\infty x^i = \frac{x}{1-x}  \, .
\end{equation}

Considering the sector as composed from an infinite set of fields with scaling dimensions $\Delta_i = i$, we obtain the following Molien-Weyl formula from Eq.\ \eqref{eq:Molien_Weyl_formula}:
\begin{equation} \label{eq:Molien_Weyl_sl(2)}
\begin{split}
	\mathcal{Z}_N^{\mathfrak{sl}(2)}(x) &= \qty(\prod_{i=1}^\infty\frac{1}{1-x^i})^N  
	 \frac{1}{(2\pi i)^{N-1}}\oint_{|t_1|=1} \frac{\dd t_1}{t_1} \cdots \oint_{|t_{N-1}|=1} \frac{\dd t_{N-1}}{t_{N-1}}  \prod_{1\leq k\leq r\leq N-1} \frac{1-t_{k,r}}{\phi_{k,r}} \,,
\end{split}
\end{equation}
where $t_{k,r}=t_k t_{k+1}\cdots t_r$ and
\begin{equation}
	\phi_{k,r} = \prod_{i=1}^\infty(1-x^it_{k,r})(1-x^it_{k,r}^{-1}) \ .
\end{equation}

For $N=1$, the Molien-Weyl formula \eqref{eq:Molien_Weyl_sl(2)} simply reduces to
\begin{equation}
	\mathcal{Z}_{N=1}^{\mathfrak{sl}(2)}(x) = \prod_{i=1}^\infty\frac{1}{1-x^i}\,,
\end{equation}
which is an infinite product of geometric series.\footnote{The inverse of this infinite product is sometimes called Euler function, and it can equally be expressed in terms of the Dedekin $\eta$ function $\eta(\tau)=q^{\frac{1}{24}}\prod_{n=1}^\infty(1-q^i)$ with $q=e^{2\pi i \tau}$.}
Under temperature inversion $T\to-T$, this expression transforms to $ix^{-\frac{1}{12}}\mathcal{Z}_{N=1}^{\mathfrak{sl}(2)}(x)$; it is thus invariant when including $x^{-\frac{1}{24}}$ as Casimir energy.%
\footnote{Concretely, 
\begin{equation}
 \mathcal{Z}_{N=1}^{\mathfrak{sl}(2)}(1/x)=\prod_{i=1}^\infty\frac{1}{1-x^{-i}}=\prod_{i=1}^\infty\frac{1}{(-1)}\prod_{i=1}^\infty x^i\prod_{i=1}^\infty\frac{1}{1-x^{i}}=\frac{x^{\zeta(-1)}}{(-1)^{\zeta(0)}}\mathcal{Z}_{N=1}^{\mathfrak{sl}(2)}(x)=ix^{-\frac{1}{12}}\mathcal{Z}_{N=1}^{\mathfrak{sl}(2)}(x)\,,
\end{equation}
where we have used $\zeta$ function regularization following Ref.\ \cite{Basar:2014mha}.
}

For $N=2$, we have
\begin{equation}
	\mathcal{Z}_{N=2}^{\mathfrak{sl}(2)}(x) = \qty(\prod_{i=1}^\infty\frac{1}{1-x^i})^2 \frac{1}{2\pi i} \oint_{|t_1|=1} \frac{\dd t_1}{t_1} \frac{1-t_1}{\phi_{1,1}} \,,
\end{equation}
where 
\begin{equation}
 \phi_{1,1}=\prod_{i=1}^\infty(1-x^it_1)(1-x^it_1^{-1})\,.
\end{equation}
This integral can be evaluated using the residue theorem by summing over all residues within the unit circle. 
Since $\phi_{1,1}$ diverges more rapidly than linearly at $t_1=0$, there is no pole at $t_1=0$ and the relevant poles come from the factor $\phi_{1,1}$ in the denominator.
Since $\abs{x} < 1$, only the poles at $t_1 = x^i$ are within the unit circle. Consider first one of the poles at $t_1 = x^j$, where the residue is
\begin{equation}
\begin{split}
	\Res_{t_1=x^j}\qty[\frac{1-t_1}{t_1}\frac{1}{\prod_{i=1}^\infty(1-x^it_1)(1-x^it_1^{-1})}]
	&= (1-x^j)\frac{1}{\prod_{i=1}^\infty(1-x^ix^j)}\frac{1}{\prod_{\substack{i=1\\i\neq j}}^\infty(1-x^ix^{-j})}
	\,.
\end{split}
\end{equation}
Next, we split $\prod_{\substack{i=1\\i\neq j}}^\infty(1-x^ix^{-j})=\prod_{i=j+1}^\infty(1-x^i)\prod_{i=1-j}^{-1}(1-x^i)$.
Since 
\begin{equation}
\prod_{i=1-j}^{-1}(1-x^i) = \prod_{i=1}^{j-1}(1-x^i)(-x^{-i}) = \frac{(-1)^{j-1}}{x^{j(j-1)/2}}\prod_{i=1}^{j-1}(1-x^i)\,,
\end{equation}
we can simplify the expression by collecting terms:
\begin{equation}
	\Res_{t_1=x^j}\qty[\cdots] = \frac{(-1)^{j-1} x^{\frac{j(j-1)}{2}} (1-x^j)^2}{\prod_{i=1}^\infty(1-x^i)^2}\,.
\end{equation}
Summing over all residues in $t_1=x^j$ for $j\in \mathbb{N}$, we obtain
\begin{equation}
\label{eq: sl2 N=2}
	\mathcal{Z}_{N=2}^{\mathfrak{sl}(2)}(x) = \frac{\sum_{i=1}^\infty (-1)^{i-1} x^{\frac{i(i-1)}{2}}(1-x^i)^2}{\prod_{i=1}^\infty (1-x^i)^4}\,.
\end{equation}
Note that the numerator is now an infinite sum instead of a polynomial as in the compact $\mathfrak{su}(2)$ and $\mathfrak{su}(2|3)$ sectors.%
\footnote{The numerator can in fact be written in terms of so-called partial theta functions.}
Each term in the sum in $\mathcal{Z}_{N=2}^{\mathfrak{sl}(2)}$ transforms with a different power of $x^{i(i+1)}$ under temperature reflection $T\to-T$. Thus, temperature reflection symmetry, at least in the usual sense, seems to be absent.
We can calculate the zeros of the $N=2$ partition function \eqref{eq: sl2 N=2} by truncating the infinite sum at some finite value $M$, observing a stabilization in the distribution of zeros as we increase $M$. 
In order to observe Lee-Yang behavior also in the $\mathfrak{sl}(2)$ sector, we would need to evaluate the Molien-Weyl formula \eqref{eq:Molien_Weyl_sl(2)} at $N>2$; this is in principle possible, but quickly becomes quite tedious.  We leave this for future work.

\subsection{\texorpdfstring{$\mathfrak{psu}(1,1|2)$}{psu(1,1|2)} sector}
\label{subsec: psu(112)}

The $\mathfrak{psu}(1,1|2)$ sector is built from two complex scalars $X=\phi_1+i\phi_4$ and $Y=\phi_2+i\phi_5$, two fermions $\psi^{\alpha=1}_{4}$ and $\bar{\psi}_{3}^{\dot\alpha=1}$ as well as a single light-like covariant derivative $\mathcal{D}^{11}=\mathcal{D}^{\mu}\sigma_{\mu}^{\alpha=1,\dot\alpha=1}$ that can act on any of them. This field content results in the single-particle partition functions
\begin{equation}
	z_B(x)  = 2\sum_{i=1}^\infty x^{i}= \frac{2x}{1-x} \ , \ \ z_F(x)  = 2\sum_{i=1}^\infty x^{i+1/2}= \frac{2x^{3/2}}{1-x} \ ,
\end{equation}
where $x=e^{-1/T}$.

Using Eq.\ \eqref{eq:Molien_Weyl_formula}, the Molien-Weyl formula for this sector takes the form
\begin{equation} \label{eq:Molien_Weyl_psu(1,1|2)}
\begin{split}
	\mathcal{Z}_N^{\mathfrak{psu}(1,1|2)}(x) &= \qty(\prod_{i=1}^\infty\frac{(1+x^{i+1/2})^{2N}}{(1-x^{i})^{2N}}) \\
	 &\phaneq\times\frac{1}{(2\pi i)^{N-1}}\oint_{|t_1|=1} \frac{\dd t_1}{t_1} \cdots \oint_{|t_{N-1}|=1} \frac{\dd t_{N-1}}{t_{N-1}}  \prod_{1\leq k\leq r\leq N-1} \frac{1-t_{k,r}}{\phi_{k,r}} \,,
\end{split}
\end{equation}
where $t_{k,r}=t_k t_{k+1}\cdots t_r$ and
\begin{equation}
	\phi_{k,r} = \prod_{i=1}^\infty \frac{(1-x^{i}t_{k,r})^2(1-x^{i}t_{k,r}^{-1})^2}{(1+x^{i+1/2}t_{k,r})^2(1+x^{i+1/2}t_{k,r}^{-1})^2} \, .
\end{equation}

For $N=1$, the Molien-Weyl formula yields
\begin{equation}
\mathcal{Z}_{N=1}^{\mathfrak{psu}(1,1|2)}(x) = \prod_{i=1}^\infty\frac{(1+x^{i+1/2})^{2}}{(1-x^{i})^2}\,.
\end{equation}

For $N=2$, we have
\begin{equation}
\begin{split}
	\mathcal{Z}_{N=2}^{\mathfrak{psu}(1,1|2)} &= \qty(\prod_{i=1}^\infty\frac{(1+x^{i+1/2})^{4}}{(1-x^{i})^4})  
	 \frac{1}{2\pi i}\oint_{|t_1|=1} \dd t_1 \frac{1-t_1}{t_1} \prod_{i=1}^\infty \frac{(1+x^{i+1/2}t_1)^2(1+x^{i+1/2}t_1^{-1})^2}{(1-x^{i}t_1)^2(1-x^{i}t_1^{-1})^2} .
\end{split}
\end{equation}
We can evaluate this integral using residue theory, similar to what was done in Section \ref{subsec: sl(2)}. The full calculation is shown in Appendix \ref{app:psu(112)_N2}. The result is the following exact partition function for the $\mathfrak{psu}(1,1|2)$ sector with $N=2$:
\begin{equation}
\label{eq: psu112 result for N=2}
\begin{split}
	\mathcal{Z}_{N=2}^{\mathfrak{psu}(1,1|2)} &= \qty(\prod_{i=1}^\infty\frac{(1+x^{i+1/2})^{8}}{(1-x^{i})^8}) (1+x^{1/2})^4 \\
	&\!\!\phaneq\times \sum_{j=1}^\infty \frac{x^{(j-1)}(1-x^j)^2 (1-x^{j-1/2}-4x^j-x^{j+1/2}-2x^{2j-1/2}-3x^{2j}-2x^{2j+1/2})}{(1+x^{j-1/2})^3(1+x^{j+1/2})^3}\,\!. 
\end{split}
\end{equation}

\subsection{Full theory}
\label{subsec: psu(224)}

The symmetry algebra of the full $\mathcal{N}=4$ SYM theory is $\mathfrak{psu}(2,2|4)$. The field content of the theory is built from six real scalars, 16 fermions and four covariant derivatives. However, terms corresponding to the equations of motion, the Bianchi identities and their derivatives have to be subtracted. The single-particle partition functions are then given by \cite{Aharony:2003sx}
\begin{equation}
\begin{split}
	z_B(x) &= \frac{6x+6x^2-14x^3+2x^4}{(1-x)^4} = \frac{6x+12x^2-2x^3}{(1-x)^3} = \sum_{i=1}^\infty (8i^2-2) x^{i} \ , \\
	z_F(x) &= \frac{16x^{3/2}-16x^{5/2}}{(1-x)^4} = \frac{16x^{3/2}}{(1-x)^3} = 8\sum_{i=1}^\infty i(i+1) x^{i+1/2} \ ,
\end{split}
\end{equation}
where $x=e^{-1/T}$. 

Using Eq.\ \eqref{eq:Molien_Weyl_formula}, the Molien-Weyl formula for the full $\mathcal{N}=4$ SYM theory is
\begin{equation}\label{eq:Molien_Weyl_psu224_v0}
\begin{split}
	\mathcal{Z}_N^{\mathfrak{psu}(2,2|4)}(x) & =\prod_{i=1}^\infty\qty(\frac{(1+x^{i+1/2})^{8i(i+1)}}{(1-x^{i})^{8i^2-2}})^{N}\\
	 &\phaneq\times\frac{1}{(2\pi i)^{N-1}}\oint_{|t_1|=1} \frac{\dd t_1}{t_1} \cdots \oint_{|t_{N-1}|=1} \frac{\dd t_{N-1}}{t_{N-1}}  \prod_{1\leq k\leq r\leq N-1} \frac{1-t_{k,r}}{\phi_{k,r}} \,,
\end{split}
\end{equation}
where $t_{k,r}=t_k t_{k+1}\cdots t_r$ and
\begin{equation}
	\phi_{k,r} = \prod_{i=1}^\infty\frac{(1-x^it_{k,r})^{8i^2-2}(1-x^it^{-1}_{k,r})^{8i^2-2}}{(1+x^{i+1/2}t_{k,r})^{8i(i+1)}(1+x^{i+1/2}t^{-1}_{k,r})^{8i(i+1)}}\,.  
\end{equation}

For $N=1$, this formula yields
\begin{equation}\label{eq:Z_psu224_N=1}
\begin{split}
	\mathcal{Z}_{N=1}^{\mathfrak{psu}(2,2|4)}(x) = \prod_{i=1}^\infty\frac{(1+x^{i+1/2})^{8i(i+1)}}{(1-x^{i})^{8i^2-2}}\,.
\end{split}
\end{equation}
While this integral can also be evaluated at larger $N$ via residues, this becomes quite tedious due to the higher order of the poles; we leave this for future work.

\section{Conclusion and outlook}
\label{sec: conclusion}

In this paper, we have studied the partition function of free $\mathcal{N}=4$ SYM theory on $\mathbb{R}\times S^3$ with gauge group $\text{U}(N)$ at finite $N$.

We obtained closed expressions for the partition function at specific values of $N$ via a Molien-Weyl formula,
 a (contour) integral over $(S^1)^{N-1}$ that can be done via residues.
We have explicitly evaluated this integral for reasonably high values of $N$ 
 in the $\mathfrak{su}(2)$ and $\mathfrak{su}(2|3)$ sector.
Moreover, as a proof of principle, we have evaluated the contour integrals for $N=2$ also in larger, non-compact sectors, namely the $\mathfrak{sl}(2)$ and $\mathfrak{psu}(1,1|2)$ sector.

In the $\mathfrak{su}(q)$ sector, the free partition function is given by a rational function in $x=e^{-1/T}$. Its numerator was conjectured to be palindromic \cite{Harmark:2014mpa}. This property can be rigorously proven via the Molien-Weyl formula \cite{teranishi1986}. Including also a Casimir energy, the partition function then becomes invariant under $T\to-T$. This symmetry has been identified and studied in a number of other physical systems in Refs.\ \cite{Basar:2014mha,McGady:2017rzv,McGady:2018rmo}.
In the higher-rank sectors including fermions that we considered, this symmetry is absent, though.

The partition function allows us to study the thermodynamic behavior of $\mathcal{N}=4$ SYM theory on $\mathbb{R}\times S^3$.
While this theory is well known to exhibit Hagedorn behavior at infinite $N$ \cite{Sundborg:1999ue}, we find that the Hagedorn behavior is replaced by Lee-Yang behavior at large but finite $N$.
Lee-Yang behavior \cite{Yang:1952be} was originally found in the description of mesoscopic systems; it means that the zeros of the partition function in the complex $x=e^{-1/T}$ plane condense in arcs that pinch the real line at the temperature of the phase transition.
Concretely, we find strong evidence for this behavior in the $\mathfrak{su}(2)$ and $\mathfrak{su}(2|3)$ sector, where the zeros appear to condense in arcs that pinch the real temperature axis at the temperature of the confinement-deconfinement transition $T_c$. 
Since the Hagedorn behavior at infinite $N$ is present in all (non-trivial) sectors including the full theory, we expect the same of Lee-Yang behavior.\footnote{It might be interesting the evaluate the corresponding Molien-Weyl formulas in Subsections \ref{subsec: sl(2)}-\ref{subsec: psu(224)} at higher $N$ to confirm this expectation.}
Thus, our findings strongly suggest that the confinement-deconfinement transition in the full $\mathcal{N}=4$ SYM theory is of Lee-Yang type as well. (While we have compelling evidence, it would be interesting to rigorously prove this.)
Via the AdS/CFT dictionary, the confinement-deconfinement transition in $\mathcal{N}=4$ SYM theory on $\mathbb{R}\times S^3$ is conjectured to be dual to the Hawking-Page transition of type IIB superstring theory on AdS${}_5\times S^5$ \cite{Witten:1998zw}. Our results thus indicate that also the Hawking-Page transition  of type IIB superstring theory on AdS${}_5\times S^5$ is of Lee-Yang type.
This is reminiscent of the results by Maloney and Witten \cite{Maloney:2007ud}, who found that the Hawking-Page transition in 3D gravity is of Lee-Yang type.

Our findings open up many interesting directions for further research.
First among these is the extension to loop level.
In the $\mathfrak{su}(2)$ sector, the action of the one-loop (and higher-loop) dilatation operator at finite $N$ can be expressed in a basis of so-called restricted Schur operators \cite{Bhattacharyya:2008xy,DeComarmond:2010ie,deMelloKoch:2012sv}. This makes it possible to determine loop corrections to the partition function (in this sector) at higher loop orders \cite{KWinprogress}.

A second direction concerns the exact calculation of the confinement-deconfinement temperature.
In the cases we studied in detail, it could be seen that the arcs of zeros of the partition function pinch the real temperature axis roughly at the known Hagedorn temperature, which is expected \cite{Aharony:2003sx} to coincides with the confinement-deconfinement temperature in the free theory. However, for fixed finite $N$, the zero closest to the real axis is still a finite distance away from the real axis.
In order to determine the exact value of the confinement-deconfinement temperature in cases where it does not coincide with the Hagedorn temperature, i.e.\ at higher loop level, it is important to be able to determine the confinement-deconfinement temperature from the asymptotic position of the closest zero to the real axis at large $N$.

Third, it would be very interesting to determine the confinement-deconfinement temperature at any value of $\lambda$ via integrability, similar to what was done for the Hagedorn temperature in Refs.\ \cite{Harmark:2017yrv,Harmark:2018red}.
For the application of integrability to the Hagedorn temperature \cite{Harmark:2017yrv,Harmark:2018red}, it was crucial to determine the pole of the partition function without needing to calculate the whole partition function. Thus, it is also likely that an application of integrability to the confinement-deconfinement temperature requires us to be able to determine the asymptotic position of the closest zero without needing to know the full partition function.

\begin{acknowledgments}
We thank 
 Robert de Mello Koch,
 Troels Harmark,
 David McGady,
 and Bo Sundborg
for very useful discussions,
David McGady for introducing the concept of Lee-Yang behavior to us,
Robert de Mello Koch,
 Troels Harmark,
 and David McGady
 for comments on the manuscript
as well as 
 Dragomir Djokovic,
and Bernd Sturmfels
for communication.
M.W.\ was supported in part by the ERC starting grant 757978 and the research grant 00015369 from Villum Fonden.
Moreover, the work of M.W.\ and A.T.K.\ was supported by the research grant 00025445 from Villum Fonden.

\end{acknowledgments}

\appendix
\section{Derivation of the character formula}
\label{app:deriv_character_formula}

In this appendix, we review the derivation of the character formula \eqref{eq:generating_function} for the partition function of free Yang-Mills theory on a compact space $\mathbb{R} \cross S^3$. We closely follow the approach outlined in Refs.\ \cite{Aharony:2003sx,Dutta:2007ws}, see also Ref.\ \cite{Skagerstam:1983gv}. 

As we are considering the theory on a compact space, the allowed states must be singlets of the gauge group. In general, we will have a number of bosonic fields with energy $E_i$, and a number of fermionic fields with energy $E_{i'}'$. We assume all fields to be in the adjoint representation of the gauge group $\text{U}(N)$. The exact partition function can then be written as
\begin{equation}
	\mathcal{Z}(\beta) = \sum_{n_1=0}^\infty x^{n_1E_1} \sum_{n_2=0}^\infty x^{n_2E_2} \dotsm \sum_{n_1'=0}^\infty x^{n_1'E_1'} \sum_{n_2'=0}^\infty x^{n_2'E_2'} \dotsm \times d(n_1,n_2,...,n_1',n_2',...) \ ,
\end{equation}
where $x=e^{-1/T}=e^{-\beta}$ and $d(n_1,n_2,...,n_1',n_2',...)$ counts the number of singlets in the product representation $\sym^{n_1}_{adj}\otimes \sym^{n_2}_{adj}\otimes \dotsm \otimes \anti^{n_1'}_{adj}\otimes \anti^{n_2'}_{adj}\otimes \dotsm $.
This factor can be calculated by integrating the characters of the representations over the group manifold, using the Haar measure $[\dd U]$. This feature stems from the orthogonality of the characters of irreducible representations and is described in detail in Appendix A of Ref.\ \cite{Aharony:2003sx}.
Thus, we obtain
\begin{equation}
	\mathcal{Z}(\beta) = \int_{\text{U}(N)}\left[ \dd U\right] \prod_i \Bigl\{ \sum_{n_i=0}^\infty x^{n_iE_i} \chi_{\sym^{n_i}_{adj}}(U) \Bigr\}   \prod_{i'} \qty{ \sum_{n'_{i'}=0}^\infty x^{n'_{i'}E'_{i'}} \chi_{\anti^{n'_{i'}}_{adj}}(U) } \ .
\end{equation}
To continue, we can utilize the following properties of group characters:
\begin{equation}
\begin{split}
	\sum_{n=0}^\infty x^n \chi_{\sym^n_R}(U) &= \exp\qty{\sum_{m=1}^\infty \frac{1}{m} x^m \chi_R (U^m) }  , \\
	\sum_{n=0}^\infty x^n \chi_{\anti^n_R}(U) &= \exp\qty{\sum_{m=1}^\infty \frac{(-1)^{m+1}}{m} x^m \chi_R (U^m) }  ,
\end{split}
\end{equation}
where $R$ can be any representation. 
If we choose $R=adj$, this yields 
\begin{equation} \label{eq:derivation_step4}
	\mathcal{Z}(\beta) = \int_{\text{U}(N)}\left[ \dd U\right]  \prod_{i}\exp \qty{ \sum_{m=1}^\infty \frac{1}{m} x^{mE_i} \chi_{adj}(U^m) }   \prod_{i'}\exp \qty{ \sum_{m=1}^\infty \frac{(-1)^{m+1}}{m} x^{mE_{i'}'} \chi_{adj}(U^m) }  .
\end{equation}
We can now identify the bosonic and fermionic single-particle partition functions
\begin{equation}
	z_B(\beta) = \sum_i x^{E_i}, \qquad z_F(\beta) = \sum_{i'} x^{E_{i'}'}  .
\end{equation}
Together with $\chi_{adj}(U) = \tr U \tr U^\dag$, we obtain a compact expression:
\begin{equation}
	\mathcal{Z}(\beta) = \int_{\text{U}(N)}\left[ \dd U\right] \exp(\sum_{j=1}^{\infty} \frac{z(j\beta)}{j}\tr U^j \tr\qty(U^\dag)^j)  ,
\end{equation}
where $z(j\beta) \equiv z_B(j\beta) - (-1)^j z_F(j\beta)$ has been introduced to simplify the notation.

For the next part of the derivation, we outline the method of Ref.\ \cite{Dutta:2007ws}. Expanding the exponential, we write
\begin{equation}
\begin{split}
	\mathcal{Z}(\beta) &= \int_{\text{U}(N)}\left[ \dd U\right] \prod_{j=1}^\infty \exp(\frac{z(j\beta)}{j}\tr U^j \tr(U^\dag)^j) \\
	&= \int_{\text{U}(N)}\left[ \dd U\right] \prod_{j=1}^\infty \sum_{k_j=0}^\infty \frac{1}{k_j!}\left( \frac{z(j\beta)}{j}\tr U^j \tr(U^\dag)^j \right)^{k_j}  .
\end{split}
\end{equation}
The product of sums over $k_j$ can be replaced by a total sum over all integer-valued vectors $\va*{k}$ in an infinite dimensional configuration space:
\begin{equation}
	\sum_{\va*{k}} \prod_{j=1}^\infty \frac{1}{k_j!}\left( \frac{z(j\beta)}{j}\tr U^j \tr(U^\dag)^j \right)^{k_j}
	= \sum_{\va*{k}} \prod_{j=1}^\infty \frac{z(j\beta)^{k_j}}{k_j! j^{k_j}} \Upsilon_{\va*{k}}(U) \Upsilon_{\va*{k}}(U^\dag) \,,
\end{equation}
where we defined
\begin{equation}
	\Upsilon_{\va*{k}}(U) = \prod_{j=1}^\infty ( \tr U^j )^{k_j} .
\end{equation}

If we let $k_j$ count the number of rows with length $j$, we can identify $\va*{k}$ with a Young tableau $k$ containing $n\equiv\sum_{j=1}^\infty j k_j$ boxes, and thus with an integer partition of $n$. With this, we can use Frobenius' formula to rewrite $\Upsilon_{\va*{k}}$ in terms of group characters as done in Ref.\ \cite{Dutta:2007ws}:
\begin{equation}
	\Upsilon_{\va*{k}}(U) = \sum_R \chi_R(k) \tr_R U \, ,
\end{equation}
where the sum is over all representations $R$ of $\text{U}(N)$. Now integrating over the Haar measure using the orthogonality of representations of $\text{U}(N)$,
\begin{equation}
	\int_{\text{U}(N)}\left[ \dd U\right] \tr_R (U) \tr_{R'} (U^\dag) = \delta_{RR'} \, ,
\end{equation}
we find
\begin{equation}
	\mathcal{Z}(\beta) = \sum_k \prod_{j=1}^\infty \frac{z(j\beta)^{k_j}}{k_j! j^{k_j}} \sum_R \abs{\chi_R(k)}^2 \, .
\end{equation}
It is convenient to group the sum over Young tableaux $k$ by the total number of boxes $n$. The representations $R$ can also be categorized as a set of Young tableaux with at most $N$ rows. We then arrive at what we call the `character formula':
\begin{equation}
	\mathcal{Z}(\beta) = 1 + \sum_{n=1}^\infty \sum_{k\vdash n} \sum_{r\vdash n} \prod_{j=1}^n \frac{z(j\beta)^{k_j}}{k_j! j^{k_j}} \abs{\chi_r(k)}^2 \ .
\end{equation}
In this formula, it is very clear to see the effect of a finite $N$: the number of rows in the representation $R$ is restricted to be no larger than $N$.

Chemical potentials can be included by replacing $E \to E -\sum_{i=1}^3 \Omega_i J_i- \sum_{a=1}^2 \Omega_{a+3} S_a$ throughout the derivation.

\section{Derivation of the Molien-Weyl formula} \label{app:deriv_Molien_Weyl_formula}

In this appendix, we review the derivation of the Molien-Weyl formula \eqref{eq:Molien_Weyl_formula} for the free partition function with general field content, cf.\ Ref.\ \cite{Dolan:2007rq}.

We take our starting point in Eq.\ \eqref{eq:derivation_step4}. We can rewrite the adjoint characters in terms of traces over the fundamental representation as $\chi_{adj}(U) = \tr U \tr U^\dag$. In the fundamental representation of $\text{U}(N)$, the group elements are $N\times N$-valued matrices. We let $\varepsilon_j$ denote the $N$ eigenvalues of $U$ with $\abs{\varepsilon_j}=1$, finding
\begin{equation}
	\chi_{adj}(U^m) = \tr U^m \tr U^{\dag m} = \qty(\sum_{r=1}^N \varepsilon_r^m) \qty(\sum_{k=1}^N \varepsilon_k^{-m}) = \sum_{k,r=1}^N \qty(\frac{\varepsilon_r}{\varepsilon_k})^m\,.
\end{equation}
Inserting this into Eq.\ \eqref{eq:derivation_step4}, we have
\begin{equation}
\begin{split}
	\mathcal{Z}(\beta) &= \int_{\text{U}(N)}\!\left[ \dd U\right] \prod_{k,r=1}^N \prod_{i}\exp \qty{ \sum_{m=1}^\infty \frac{1}{m} x^{mE_i} \qty(\frac{\varepsilon_r}{\varepsilon_k})^m }   \prod_{i'}\exp \qty{ \sum_{m=1}^\infty \frac{(-1)^{m+1}}{m} x^{mE_{i'}'} \qty(\frac{\varepsilon_r}{\varepsilon_k})^m }\\
	&= \int_{\text{U}(N)}\left[ \dd U\right] \prod_{k,r=1}^N \frac{\prod_{i'} \qty(1+x^{E_{i'}'}\frac{\varepsilon_r}{\varepsilon_k})}{\prod_i \qty(1-x^{E_i}\frac{\varepsilon_r}{\varepsilon_k})}\\
	&= \qty(\frac{\prod_{i'} (1+x^{E_{i'}'})}{\prod_i (1-x^{E_i})})^N   \int_{\text{U}(N)}\left[ \dd U\right] \prod_{1\leq k<r \leq N} \frac{\prod_{i'} \qty(1+x^{E_{i'}'}\frac{\varepsilon_r}{\varepsilon_k})\qty(1+x^{E_{i'}'}\frac{\varepsilon_k}{\varepsilon_r})}{\prod_i \qty(1-x^{E_i}\frac{\varepsilon_r}{\varepsilon_k})\qty(1-x^{E_i}\frac{\varepsilon_k}{\varepsilon_r})} \ .
\end{split}
\end{equation}
In the second line we used the fact that $-\log(1-x) = \sum_{m=1}^\infty \frac{1}{m}x^m$, and in the third line we split the product into three parts with $k=r$, $k<r$ and $k>r$, respectively.

The Haar measure can be replaced by an integral over the eigenvalues of the $\text{U}(N)$ group element $U$, on the unit circle $\abs{\varepsilon_j}=1$ \cite{Weyl:1939}:
\begin{equation}
	\int_{\text{U}(N)}\left[ \dd U\right] = \frac{1}{N! (2\pi i)^N} \oint \prod_{j=1}^N \frac{\dd \varepsilon_j}{\varepsilon_j} \ \Delta\bar{\Delta} \ ,
\end{equation}
where the Vandermonde determinant is given by 
\begin{equation}
 \Delta = \prod_{k<r} (\varepsilon_r-\varepsilon_k)\,\qquad\bar{\Delta} = \prod_{k<r} (\varepsilon_r^{-1}-\varepsilon_k^{-1})\,.
\end{equation}
In conclusion, we can compute the partition function of any field content by the generalized Molien-Weyl formula:
\begin{equation}\label{eq:Molien_Weyl_alternative}
	\mathcal{Z}(x) = \qty(\mathcal{Z}_{N=1}(x))^N   \frac{1}{N! (2\pi i)^N} \oint \prod_{j=1}^N \frac{\dd \varepsilon_j}{\varepsilon_j} \ \Delta\bar{\Delta}  \prod_{1\leq k<r \leq N} \frac{1}{\phi_{k,r}}\ ,
\end{equation}
where we have introduced
\begin{align}
	\mathcal{Z}_{N=1}(x) &= \frac{\prod_{i'} (1+x^{E_{i'}'})}{\prod_i (1-x^{E_i})} \, ,\\
	\phi_{k,r} &=  \frac{\prod_i \qty(1-x^{E_i}\frac{\varepsilon_r}{\varepsilon_k})\qty(1-x^{E_i}\frac{\varepsilon_k}{\varepsilon_r})}{\prod_{i'} \qty(1+x^{E_{i'}'}\frac{\varepsilon_r}{\varepsilon_k})\qty(1+x^{E_{i'}'}\frac{\varepsilon_k}{\varepsilon_r})} \, .
\end{align}

We can obtain another form of the Molien-Weyl formula by utilizing the symmetry under a permutation of the $N$ eigenvalues $\varepsilon_i$.
First, notice that the measure $\prod_{j=1}^N \frac{\dd \varepsilon_j}{\varepsilon_j}$, $\Delta\bar{\Delta}$ and $\prod_{k<r} \phi_{k,r}(x)$ are all invariant under a permutation of the $\varepsilon_i$ while $\Delta$ and $\bar{\Delta}$ are each invariant up to the sign of the permutation, $\sgn(\sigma)$.
By its definition, the Vandermonde determinant can also be written as 
\begin{equation}
\begin{split}
	\Delta = \prod_{k<r} (\varepsilon_r-\varepsilon_k)
	= \sum_{\sigma\in S_N} \sgn(\sigma)\ \varepsilon_{\sigma(1)}^0\varepsilon_{\sigma(2)}^1 \cdots \varepsilon_{\sigma(N)}^{N-1}\,.
\end{split}
\end{equation}
Using the freedom to relabel the $\varepsilon_i$, we can permute the $\varepsilon_i$ in each term with the inverse permutation $\sigma^{-1}$ to make them all be of the form $\varepsilon_{1}^0\varepsilon_{2}^1 \cdots \varepsilon_{N}^{N-1}$. When doing so, $\bar{\Delta}$ picks up a factor of $\sgn(\sigma^{-1})$ which cancels the sign factor in $\Delta$, and we are left with
\begin{equation}
	\Delta\bar{\Delta} \to N! \varepsilon_{1}^0\varepsilon_{2}^1 \cdots \varepsilon_{N}^{N-1} \bar{\Delta} = N! \prod_{k<r} \qty(1-\frac{\varepsilon_r}{\varepsilon_k})\,.
\end{equation}
We thus obtain the alternative Molien-Weyl formula
\begin{equation}
\mathcal{Z}(x) = \qty(\mathcal{Z}_{N=1}(x))^N   \frac{1}{(2\pi i)^N} \oint \prod_{j=1}^N \frac{\dd \varepsilon_j}{\varepsilon_j} \ \prod_{1\leq k<r \leq N} \qty(1-\frac{\varepsilon_r}{\varepsilon_k}) \frac{1}{\phi_{k,r}}\,.
\end{equation}

Yet another form of the Molien-Weyl formula can be obtained by a change of variables: $\varepsilon_j = t_1\cdots t_j$. With this identification, we have
\begin{equation}
	\prod_{1\leq k<r \leq N} \left(1 \pm x\frac{\varepsilon_r}{\varepsilon_k}\right)\left(1 \pm x\frac{\varepsilon_k}{\varepsilon_r}\right) = \prod_{2\leq k\leq r \leq N} (1 \pm x t_{k,r})(1 \pm xt_{k,r}^{-1}) \ ,
\end{equation}
and
\begin{equation}
	\prod_{1\leq k<r\leq N} \qty(1-\frac{\varepsilon_r}{\varepsilon_k}) = \prod_{2\leq k\leq r\leq N} (1-t_{k,r}) \ ,
\end{equation}
where $t_{k,r}=t_k t_{k+1}\cdots t_r$.
The Jacobian for the transformation is given by
\begin{equation}
	J = \det[\dv{\varepsilon_i}{t_j}] = t_1^{N-1}t_2^{N-2}\cdots t_{N-1}^1 t_N^0 = \prod_{j=1}^N \frac{\varepsilon_j}{t_j} \ .
\end{equation}
In total, we can rewrite the generalized Molien-Weyl formula as 
\begin{equation}
	\mathcal{Z}(x) = \qty(\mathcal{Z}_{N=1}(x))^N
	\frac{1}{(2\pi i)^N} \oint \prod_{j=1}^N \frac{\dd t_j}{t_j}
	\prod_{1\leq k\leq r \leq N-1}  \frac{1-t_{k,r}}{\phi_{k,r}} \ ,
\end{equation}
where we have relabeled the integration variables $t_N\to t_{N-1}\to \dotsc \to t_1\to t_N$
and
\begin{align}
	\mathcal{Z}_{N=1}(x) &= \frac{\prod_{i'} (1+x^{E_{i'}'})}{\prod_i \qty(1-x^{E_i})} \ ,
\\
	\phi_{k,r} &= \frac{\prod_i (1-x^{E_i}t_{k,r})(1-x^{E_i}t_{k,r}^{-1})}{\prod_{i'} (1+x^{E_{i'}'}t_{k,r})(1+x^{E_{i'}'}t_{k,r}^{-1})} \ .
\end{align}
Note that $t_N$ only occurs once in the integrand as $1/t_N$. Integrating $t_N$ then simply gives a factor of $2\pi i$, and we have
\begin{equation}
	\mathcal{Z}(x) = \qty(\mathcal{Z}_{N=1}(x))^N
	\frac{1}{(2\pi i)^{N-1}} \oint_{|t_1|=1} \frac{\dd t_1}{t_1} \cdots \oint_{|t_{N-1}|=1} \frac{\dd t_{N-1}}{t_{N-1}}   \prod_{1\leq k\leq r \leq N-1}  \frac{1-t_{k,r}}{\phi_{k,r}} \ ,
\end{equation}
as claimed in Eq.\ \eqref{eq:Molien_Weyl_formula} of the main text.

Chemical potentials can be included by replacing $E \to E -\sum_{i=1}^3 \Omega_i J_i- \sum_{a=1}^2 \Omega_{a+3} S_a$ throughout the derivation.

\section{Calculation of the partition function in the \texorpdfstring{$\mathfrak{psu}(1,1|2)$}{psu(1,1|2)} sector at \texorpdfstring{$N=2$}{N=2}}
\label{app:psu(112)_N2}

In this appendix, we provide details on the calculation of the partition function in the $\mathfrak{psu}(1,1|2)$ sector at $N=2$, the result of which is given in Eq.\ \eqref{eq: psu112 result for N=2}.

Our starting point is the Molien-Weyl formula \eqref{eq:Molien_Weyl_psu(1,1|2)} for this sector with $N=2$, which we quote here for convenience: 
\begin{equation} \label{eq:psu112_N=2}
\begin{split}
	\mathcal{Z}_{N=2}^{\mathfrak{psu}(1,1|2)} &= \qty(\prod_{i=1}^\infty\frac{(1+x^{i+1/2})^{4}}{(1-x^{i})^4}) 
	 \frac{1}{2\pi i}\oint_{|t_1|=1} \dd t_1 \frac{1-t_1}{t_1} \prod_{i=1}^\infty \frac{(1+x^{i+1/2}t_1)^2(1+x^{i+1/2}t_1^{-1})^2}{(1-x^{i}t_1)^2(1-x^{i}t_1^{-1})^2} .
\end{split}
\end{equation}

The relevant poles are double poles in $t_1 = x^j$ for all $j\in \mathbb{N}$.%
\footnote{%
While there is a pole at $t_1=0$, the corresponding residue is
\begin{equation}
\begin{split}
	(1-t_1) \prod_{i=1}^\infty \frac{(1+x^{i+1/2}t_1)^2(1+x^{i+1/2}t_1^{-1})^2}{(1-x^{i}t_1)^2(1-x^{i}t_1^{-1})^2}\Big|_{t_1=0}
	&=  \prod_{i=1}^\infty \frac{(t_1+x^{i+1/2})^2 t_1^2}{(t_1-x^{i})^2 t_1^2}\Big|_{t_1=0}\\
	&
	= \prod_{i=1}^\infty \frac{(x^{i+1/2})^2}{(-x^{i})^2}= \prod_{i=1}^\infty x\,.
\end{split}
\end{equation}
Since $\abs{x}<1$, this infinite product goes to $0$, and the pole at $t_1=0$ does not contribute.}
Focusing on a single value of $j$, the residue of the pole at $t_1 = x^j$ is
\begin{equation}
\begin{split}
	&\Res_{t_1=x^j}\Bigl[\frac{1-t_1}{t_1} \prod_{i=1}^\infty \frac{(1+x^{i+1/2}t_1)^2(1+x^{i+1/2}t_1^{-1})^2}{(1-x^{i}t_1)^2(1-x^{i}t_1^{-1})^2}\Bigr]\\
	&= \pdv{t_1}\Bigl[\underbrace{t_1(1-t_1)}_{\mathtt{pre}} \underbrace{\prod_{i=1}^\infty(1+x^{i+1/2}t_1)^2(1+x^{i+1/2}t_1^{-1})^2}_{\mathtt{ferm}} \underbrace{\prod_{i=1}^\infty\frac{1}{(1-x^{i}t_1)^2} \prod_{\substack{i=1\\i\neq j}}^\infty \frac{1}{(1-x^{i}t_1^{-1})^2}}_{\mathtt{scal}}\Bigr]\Big|_{t_1=x^j}.
\end{split}
\end{equation}
We now split the action of the derivative onto the prefactor $t_1(1-t_1)$, the fermionic terms in the numerator and the scalar terms in the denominator. The action on the prefactor yields
\begin{equation}
	(\partial_{t_1}\mathtt{pre})(\mathtt{ferm})(\mathtt{scal})|_{t_1=x^j}=(1-2x^j) \Gamma(x^j)\,,
\end{equation}
where 
\begin{equation}
	\Gamma(x^j) = \lim_{t_1\to x^j}  \prod_{i=1}^\infty \frac{(1+x^{i+1/2}t_1)^2(1+x^{i+1/2}t_1^{-1})^2}{(1-x^{i}t_1)^2} \prod_{\substack{i=1\\i\neq j}}^\infty \frac{1}{(1-x^{i}t_1^{-1})^2}\,.
\end{equation}
The action on the scalar terms yields
\begin{equation}
	(\mathtt{pre})(\mathtt{ferm})(\partial_{t_1}\mathtt{scal})|_{t_1=x^j}=x^j(1-x^j) \qty(\sum_{k=1}^\infty \frac{2x^k}{1-x^kx^j} - \sum_{\substack{k=1\\k\neq j}}^\infty \frac{2x^kx^{-2j}}{1-x^kx^{-j}}) \Gamma(x^j)\,.
\end{equation}
The infinite sums can be reduced in the following manner:
\begin{equation}
\begin{split}
	\sum_{k=1}^\infty \frac{x^k}{1-x^kx^j} - \sum_{\substack{k=1\\k\neq j}}^\infty \frac{x^kx^{-2j}}{1-x^kx^{-j}}
	&= \sum_{i=j+1}^\infty \frac{x^{i-j}}{1-x^i} - \sum_{i=1-j}^{-1} \frac{x^{i-j}}{1-x^i} - \sum_{i=1}^{\infty} \frac{x^{i-j}}{1-x^i}\\
	&= + \sum_{i=1}^{j-1} \frac{x^{-j}}{1-x^i} - \sum_{i=1}^{j} \frac{x^{i-j}}{1-x^i}\\
	&= \sum_{i=1}^{j-1} \frac{x^{-j}-x^{i-j}}{1-x^i} - \frac{1}{1-x^j} = x^{-j}(j-1) - \frac{1}{1-x^j}\,.
\end{split}
\end{equation}
In total, the derivative on the scalar terms contributes with
\begin{equation}
\begin{split}
	(\mathtt{pre})(\mathtt{ferm})(\partial_{t_1}\mathtt{scal})|_{t_1=x^j}&=2x^j(1-x^j) \qty(x^{-j}(j-1) - \frac{1}{1-x^j}) \Gamma(x^j)\\
	&= \qty(2(1-x^j)(j-1) - 2x^j) \Gamma(x^j)\,.
\end{split}
\end{equation}
Now we calculate the effect of the derivative on the fermionic terms:
\begin{equation}
	(\mathtt{pre})(\partial_{t_1}\mathtt{ferm})(\mathtt{scal})|_{t_1=x^j}=x^j(1-x^j) \sum_{k=1}^\infty \qty( \frac{2x^{k+1/2}}{1+x^{k+1/2}x^j} - \frac{2x^{k+1/2}x^{-2j}}{1+x^{k+1/2}x^{-j}} )  \Gamma(x^j)\,.
\end{equation}
Again, the infinite sums can be reduced:
\begin{equation}
\begin{split}
	&\sum_{k=1}^\infty \qty( \frac{x^{k+1/2}}{1+x^{k+1/2}x^j} - \frac{x^{k+1/2}x^{-2j}}{1+x^{k+1/2}x^{-j}} )\\
	= & \sum_{i=j+1}^\infty \frac{x^{i-j+1/2}}{1+x^{i+1/2}} -  \sum_{k=1}^{j-2} \frac{x^{k+1/2}x^{-2j}}{1+x^{k+1/2}x^{-j}} - \frac{x^{-j+1/2}}{1+x^{1/2}} - \frac{x^{-j-1/2}}{1+x^{-1/2}} - \sum_{i=1}^\infty \frac{x^{i-j+1/2}}{1+x^{i+1/2}} \,.
\end{split}
\end{equation}
The two terms outside the sums join to become $-x^{-j}$. 
The first sum cancels all except the first $j$ terms in the last sum. 
The second sum becomes
\begin{equation}
\begin{split}
	-\sum_{k=1}^{j-2} \frac{x^{k+1/2}x^{-2j}}{1+x^{k+1/2}x^{-j}}
	= -\sum_{k=-j+2}^{-1} \frac{x^{k-1/2}x^{-j}}{1+x^{k-1/2}}
	= -\sum_{k=-j+2}^{-1} \frac{x^{-j}}{1+x^{-k+1/2}}
	= -\sum_{i=1}^{j-2} \frac{x^{-j}}{1+x^{i+1/2}}\,.
\end{split}
\end{equation}
With these rearrangements, we obtain
\begin{equation}
\begin{split}
	-x^{-j} - \sum_{i=1}^{j} \frac{x^{i-j+1/2}}{1+x^{i+1/2}} - \sum_{i=1}^{j-2} \frac{x^{-j}}{1+x^{i+1/2}}
	= &-x^{-j} - x^{-j}(j-2) - \frac{x^{-1/2}}{1+x^{j-1/2}} - \frac{x^{+1/2}}{1+x^{j+1/2}}\\
	= &-x^{-j}\qty( (j-1) + \frac{x^{j-1/2}}{1+x^{j-1/2}} + \frac{x^{j+1/2}}{1+x^{j+1/2}} )\,.
\end{split}
\end{equation}
Thus, the total contribution from the derivative acting on the fermionic terms is
\begin{equation}
	(\mathtt{pre})(\partial_{t_1}\mathtt{ferm})(\mathtt{scal})|_{t_1=x^j}=-2(1-x^j)\qty( (j-1) + \frac{x^{j-1/2}}{1+x^{j-1/2}} + \frac{x^{j+1/2}}{1+x^{j+1/2}} )  \Gamma(x^j)\,.
\end{equation}
Adding the three contributions from the prefactor, the scalar and the fermionic terms, we find the full residue from the pole at $t_1=x^j$:
\begin{equation}\label{eq:psu112_residue}
\begin{split}
	\Res_{t_1=x^j}\qty[\cdots]&=\Bigl[ (1-2x^j) + 2(1-x^j)(j-1) - 2x^j 
	\\&\qquad- 2(1-x^j)\qty( (j-1) + \frac{x^{j-1/2}}{1+x^{j-1/2}} + \frac{x^{j+1/2}}{1+x^{j+1/2}} ) \Bigr]  \Gamma(x^j)\\
	&= \qty[1-4x^j -2(1-x^j)\qty( \frac{x^{j-1/2}}{1+x^{j-1/2}} + \frac{x^{j+1/2}}{1+x^{j+1/2}} )]  \Gamma(x^j)\\
	&= \frac{1-x^{j-1/2}-4x^j-x^{j+1/2}-2x^{2j-1/2}-3x^{2j}-2x^{2j+1/2}}{(1+x^{j-1/2})(1+x^{j+1/2})}  \Gamma(x^j)\,.
\end{split}
\end{equation}

Finally we will take a look at the infinite products in $\Gamma(x^j)$. The infinite products in the denominators can be rearranged as
\begin{equation}
	\prod_{\substack{i=1\\i\neq j}}^\infty (1-x^{i-j}) \prod_{i=1}^\infty (1-x^{i+j}) = \prod_{i=1-j}^{-1} (1-x^i)\prod_{i=1}^\infty (1-x^i) \prod_{i=1+j}^\infty (1-x^i)\,.
\end{equation}
Since $\prod_{i=1-j}^{-1}(1-x^i) = \prod_{i=1}^{j-1}(1-x^i)(-x^{-i}) = \frac{(-1)^{j-1}}{x^{j(j-1)/2}}\prod_{i=1}^{j-1}(1-x^i)$, we can simplify the expression by collecting terms:
\begin{equation}
\begin{split}
	\prod_{\substack{i=1\\i\neq j}}^\infty (1-x^{i-j}) \prod_{i=1}^\infty (1-x^{i+j}) = \frac{\prod_{i=1}^\infty (1-x^i)^2}{x^{j(j-1)/2}(1-x^j)}\,. 
\end{split}
\end{equation}
The infinite products in the numerator can be rearranged as
\begin{equation}
\begin{split}
	\prod_{i=1}^\infty (1+x^{i+1/2}x^j)(1+x^{i+1/2}x^{-j})
	&= \prod_{i=1}^\infty (1+x^{i+j+1/2})(1+x^{i-j+1/2})\\
	&= \prod_{i=1+j}^\infty (1+x^{i+1/2}) \prod_{i=1-j}^{0} (1+x^{i+1/2}) \prod_{i=1}^\infty (1+x^{i+1/2})\,.
\end{split}
\end{equation}
The middle term can be rewritten as
\begin{equation}
\begin{split}
	\prod_{i=1-j}^{0} (1+x^{i+1/2}) &= \prod_{i=0}^{j-1} (1+x^{i-1/2})x^{-i+1/2}
	 = x^{j/2-j(j-1)/2} \prod_{i=-1}^{j-2} (1+x^{i+1/2})\\
	&= x^{-j(j-2)/2} \frac{(1+x^{-1/2})(1+x^{1/2})}{(1+x^{j-1/2})(1+x^{j+1/2})} \prod_{i=1}^{j} (1+x^{i+1/2})\\
	&= x^{-j(j-2)/2 - 1/2} \frac{(1+x^{1/2})^2}{(1+x^{j-1/2})(1+x^{j+1/2})} \prod_{i=1}^{j} (1+x^{i+1/2})\\
	&= x^{-(j-1)^2/2} \frac{(1+x^{1/2})^2}{(1+x^{j-1/2})(1+x^{j+1/2})} \prod_{i=1}^{j} (1+x^{i+1/2})\,.
\end{split}
\end{equation}
We find
\begin{equation}
	\prod_{i=1}^\infty (1+x^{i+1/2}x^j)(1+x^{i+1/2}x^{-j}) = \frac{x^{-(j-1)^2/2}(1+x^{1/2})^2}{(1+x^{j-1/2})(1+x^{j+1/2})} \prod_{i=1}^\infty (1+x^{i+1/2})^2\,.
\end{equation}
In total, the $\Gamma(x^j)$ factor thus takes the form
\begin{equation}\label{eq:psu112_Gamma}
\begin{split}
	\Gamma(x^j) &= \frac{ \prod_{i=1}^\infty (1+x^{i+1/2}x^j)^2(1+x^{i+1/2}x^{-j})^2 }{ \prod_{i=1}^\infty (1-x^{i+j})^2 \prod_{\substack{i=1\\i\neq j}}^\infty (1-x^{i-j})^2 }\\
	&= \frac{x^{j(j-1)}(1-x^j)^2}{\prod_{i=1}^\infty (1-x^i)^4}  \frac{x^{-(j-1)^2} (1+x^{1/2})^4}{(1+x^{j-1/2})^2(1+x^{j+1/2})^2} \prod_{i=1}^\infty (1+x^{i+1/2})^4\\
	&= \frac{\prod_{i=1}^\infty (1+x^{i+1/2})^4}{\prod_{i=1}^\infty (1-x^i)^4}  (1+x^{1/2})^4  \frac{x^{(j-1)} (1-x^j)^2}{(1+x^{j-1/2})^2(1+x^{j+1/2})^2} \,.
\end{split}
\end{equation}

Combining Eqs.\ \eqref{eq:psu112_residue} and \eqref{eq:psu112_Gamma}, we obtain the rearranged residue of the pole at $t_1=x^j$. Summing this expression over all $j\in \mathbb{N}$ we have included all residues, and (remembering the prefactor in the first line of Eq.\ \eqref{eq:psu112_N=2}) we obtain the final partition function for the $\mathfrak{psu}(1,1|2)$ sector for $N=2$:
\begin{equation}
\label{eq: psu112 result for N=2 appendix}
\begin{split}
	\mathcal{Z}_{N=2}^{\mathfrak{psu}(1,1|2)} &= \qty(\prod_{i=1}^\infty\frac{(1+x^{i+1/2})^{8}}{(1-x^{i})^8}) (1+x^{1/2})^4 \\
	&\!\!\phaneq\times \sum_{j=1}^\infty \frac{x^{(j-1)}(1-x^j)^2 (1-x^{j-1/2}-4x^j-x^{j+1/2}-2x^{2j-1/2}-3x^{2j}-2x^{2j+1/2})}{(1+x^{j-1/2})^3(1+x^{j+1/2})^3}\,\!, 
\end{split}
\end{equation}
as claimed in Eq.\ \eqref{eq: psu112 result for N=2} of the main text.

\bibliographystyle{utphys2}
\bibliography{bibfile}

\end{document}